\documentclass[11pt,oneside,letterpaper]{article}
\usepackage[utf8]{inputenc}
\usepackage{graphicx}
\usepackage{amsmath}
\usepackage{amsthm}
\usepackage{amssymb}
\usepackage{bbm}
\usepackage{physics}
\usepackage{tensor}
\usepackage{slashed}
\usepackage[letterpaper, total={6.6in, 9in}]{geometry}
\usepackage{mathrsfs}
\usepackage[linktocpage]{hyperref}
\usepackage{setspace}
\usepackage{comment}
\usepackage{tikz}
\usepackage{array}
\usepackage{longtable}
\usepackage{mathtools}
\usepackage{afterpage}
\usepackage{authblk}

\newcommand{\be}{\begin{equation}}
\newcommand{\ee}{\end{equation}}
\newcommand{\bea}{\begin{eqnarray}}
\newcommand{\eea}{\end{eqnarray}}
\newcommand{\bel}{\begin{align}}
\newcommand{\eel}{\end{align}}
\newcommand{\bi}{\begin{itemize}}
\newcommand{\ei}{\end{itemize}}
\newcommand{\id}{{\mathbbm{1}}}

\makeatletter\@addtoreset{equation}{section}\makeatother 

\newcommand\mc[1]{\mathcal{#1}}
\newcommand\mb[1]{\mathbb{#1}}
\newcommand\calo{\mathcal{O}}

\definecolor{vert}{rgb}{0.1367 0.543 0.1367}

\DeclareMathOperator{\MCG}{MCG}
\DeclareMathOperator{\Diff}{Diff}

\DeclareMathOperator{\diag}{diag}
\newcommand\half{{1\over 2}}

\hypersetup{
    colorlinks=true,
    citecolor=blue,
    linkcolor=blue,
    filecolor=magenta,      
    urlcolor=cyan,
    pdftitle={Overleaf Example},
    pdfpagemode=FullScreen,
    }

\title{{\huge Resurgence, Conformal Blocks, and the Sum over Geometries in Quantum Gravity} \vspace{1cm}}

\author[a]{Nathan Benjamin}
\author[b]{Scott Collier}
\author[c]{Alexander Maloney}
\author[c]{Viraj Meruliya\vspace{0.8cm}}

\affil[a]{  \textit{{Walter Burke Institute for Theoretical Physics, Caltech, Pasadena, CA 91125, USA
}}\vspace{0.2cm}}
\affil[b]{\textit{{Princeton Center for Theoretical Science, Princeton University, Princeton, NJ 08544, USA }\vspace{0.2cm}}
}
\affil[c]{\textit{{Department of Physics, McGill University, Montreal, QC H3A 2T8, Canada }}}

\date{}

\begin{document}

\maketitle

\thispagestyle{empty}

\begin{center}
  \texttt{nbenjami@caltech.edu, scott.collier@princeton.edu \\ maloney@physics.mcgill.ca, viraj.meruliya@mail.mcgill.ca}
\end{center}
\vspace{1cm}

\begin{abstract}
In two dimensional conformal field theories the limit of large central charge plays the role of a semi-classical limit.  Certain universal observables, such as conformal blocks involving the exchange of the identity operator, can be expanded around this classical limit in powers of the central charge $c$.  This expansion is an asymptotic series, so -- via the same resurgence analysis familiar from quantum mechanics -- necessitates the existence of non-perturbative effects.  
In the case of identity conformal blocks, these new effects have a simple interpretation: the CFT must possess new primary operators with dimension of order the central charge.  This constrains the data of CFTs with large central charge in a way that is similar to (but distinct from) the conformal bootstrap.
We study this phenomenon in three ways:
numerically, analytically using Zamolodchikov's recursion relations, and by considering non-unitary minimal models with large (negative) central charge.
In the holographic dual to a CFT$_2$, the expansion in powers of $c$ is the perturbative loop expansion in powers of $\hbar$.  So our results imply that the graviton loop expansion is an asymptotic series, whose cure requires the inclusion of new saddle points in the gravitational path integral.  In certain cases these saddle points have a simple interpretation: they are conical excesses, particle-like states with negative mass which are not in the physical spectrum but nevertheless appear as non-manifold saddle points that control the asymptotic behaviour of the loop expansion.  
This phenomenon also has an interpretation in $SL(2,{\mathbb R})$ Chern-Simons theory, where the non-perturbative effects are associated with the non-Teichm\"uller component of the moduli space of flat connections.

\end{abstract}

\newpage
\pagenumbering{arabic}
\tableofcontents

\section{Introduction}

Many perturbative expansions in physics -- from standard time-independent perturbation theory in quantum mechanics to the loop expansion in field theory -- do not have have a finite radius of convergence.  Instead, these expansions are typically divergent asymptotic series, and non-perturbative effects must be included in order to render the theory sensible.  These non-perturbative effects often take the form of instantons: new, subleading saddle points in the path integral of the theory.  In some cases these new saddle points are physical -- such as instantons which mediate decay or tunnelling effects.  In other cases the saddle points are unphysical, possibly complex, solutions to the equations of motion which do not lie on the contour of integration for the path integral but nevertheless control the behaviour of the asymptotic expansion.  In both cases, if the perturbative expansion can be understood in sufficient detail, the behaviour of the asymptotic series can be used to extract novel information about the path intergral and the non-perturbative structure of the theory.

Our goal is to apply this logic to two dimensional conformal field theories, where the inverse of the central charge $c$ plays the role of the small perturbative parameter.  The objects that we will consider are conformal blocks which package the contribution to an observable (such as a correlation function or partition function) of an entire representation of the Virasoro algebra.
A typical example is the four point function of an operator ${\cal V}$, which can be written as a sum over intermediate states as
\be\label{4pt}
\langle {\cal V}(0) {\cal V}(x) {\cal V} (1) {\cal V}(\infty) \rangle = \left|{\cal F}_{\id}(c; x)\right|^2 + \sum_{\cal O} C_{{\cal V}{\cal V} {\cal O}}^2 \left|{\cal F}_{h_{\cal O}} (c;x)\right|^2~.
\ee
Here the identity block ${\cal F}_{\id}$ packages together all of the contributions from the identity operator and its descendants, and 
${\cal F}_{h_{\cal O}}$ similarly describes the contributions from a primary
${\cal O}$ of dimension $h_{\cal O}$ and its descendants.
The conformal blocks depend on $c$, on the cross-ratio $x$, and on the dimensions of all of the operators involved.\footnote{For brevity, we often omit the dependence on the external operator dimensions.}
In certain kinematic regimes, such as when the cross ratio $x$ is small, the exchange of the identity operator dominates.

In this paper we will be interested in the behaviour of the identity conformal block ${\cal F}_{\id}(c;x)$.
This block has a perturbative expansion at large $c$ of the form\footnote{In this expression we have written the $1/c$ expansion of the logarithm of the block, although later we will find it more convenient to work with an expansion of the block itself.  An asymptotic series in one is related to an asymptotic series in the other.}
\be\label{fexp}
{\cal F}_{\id}(c; x) \approx \exp\left\{ c f(x) + \sum_{n=0}^\infty f^{(n)}(x) c^{-n}
\right\}
\ee
The leading $f(x)$ term is known as a classical conformal block, and the subleading $f^{(c)}$ represent the ``quantum" corrections to this classical block.  Here the central charge $c$ plays the role of $\hbar^{-1}$.\footnote{In many contexts it is more natural to think of $b^{-1}$, with $c=1+6 \left(b+b^{-1}\right)^2$, as playing the role of $\hbar^{-1}$.  For our purposes either $b^{-1}$ or $c$ will do.}  Although efficient methods exist to compute conformal blocks perturbatively in $x$, at finite values of $x$ it is typically impossible to compute ${\cal F}_{\id}(c; x)$ exactly.  Even the classical block $f(x)$ is only known analytically in certain very special cases.  The subleading terms are likewise generally not known analytically.

Nevertheless, various techniques will allow us to study the nature of the series expansion in equation (\ref{fexp}).  When we do so, our conclusion will be that the expansion (\ref{fexp}) is a divergent, asymptotic series in $c^{-1}$.\footnote{This observation has been made in several other contexts; see e.g. \cite{Fitzpatrick:2016ive} for a recent discussion.}   We will describe three different methods which allow us to study this asymptotic expansion.  The first is by considering the nature of two dimensional CFTs at large {\it negative} central charge.  After all, if the series expansion (\ref{fexp}) had a finite radius of convergence around $c=\infty$, then we could freely analytically continue from large positive to large negative central charge.  But CFTs with negative $c$ are non-unitary, so this cannot be possible.  So the series expansion cannot converge.  Indeed, this is just a version of Dyson's original argument that the perturbative expansion of QED diverges \cite{Dyson:1952tj}, but with the CFT central charge playing the role of the fine structure constant.   
Two dimensional CFTs are considerably simpler than QED, however, since non-unitary but exactly solvable CFTs exist with large (negative) values of the central charge.  We will analyze their conformal blocks explicitly using the Coulomb gas representation, and prove that the resulting series is asymptotic.  

For more complicated, irrational CFTs other methods must be used to study the nature of the $1/c$ expansion. We will use Zamolodchikov's recursion relations \cite{1984CMaPh..96..419Z,1987TMP....73.1088Z}.  Although typically used to efficiently compute the terms in the $x$-expansion, they can also be used to study the structure of the $1/c$ expansion.  Indeed, one formulation of these recursion relations (the ``$c$-recursion" rather than ``$h$-recursion") is well suited to this analysis, since it determines terms in the cross-ratio expansion of conformal blocks as meromorphic functions of the central charge.  We will see that, in a certain approximation, this allows us to extract the behaviour of the $1/c$ expansion. In this approximation we can determine exactly the location of poles in the Borel plane, and so demonstrate that conformal blocks are asmpytotic series.  Finally, we will verify this behaviour using direct, numerical computation of the conformal blocks.

What are the implications of this result?  In a typical quantum mechanical system an asymptotic series indicates the existence of non-perturbative effects. 
These often take the form of new solutions to the equations of motion which are subleading compared to the solution being expanded around.\footnote{A second type of effect -- the renormalon -- is associated with the structure of the Feynman diagram expansion of field theories in either the deep IR or UV.  Renormalons do not seem to play a role in the $1/c$ expansion of two dimensional CFTs.}
More specifically, the asymptotic growth of terms the perturbative expansion is determined by the action of this new saddle point. In some cases, if the perturbative series is Borel summable, one can explicitly resum the perturbative series and infer the existence of not just one but many novel saddle points.  We will see that, at least in a certain approximation, the perturbative expansion of conformal blocks has this property: we will explicitly locate the corresponding divergences in the Borel plane.

What is the interpretation of these new saddle points?  From the point of view of our original conformal block expansion (\ref{4pt}) it is not hard to anticipate the answer: they are other conformal blocks, corresponding to the exchange of non-trivial primary operators ${\cal O}$.   In other words, the structure of the perturbative $1/c$ expansion of the identity block ${\cal F}_\id$ implies that a CFT with large central charge must include new primary states aside from the identity operator. 
In particular, these primary states must have dimension $h_{\cal O} \sim {\cal O}(c)$.  We will see this quite explicitly for minimal models at negative central charge, as well as more generally for unitary CFTs with large positive central charge.  Curiously, this result is quite similar to those which follow from bootstrap arguments (see e.g. \cite{Poland:2018epd, Hellerman:2009bu}), although the logic of the argument is entirely different.  So our asymptotic methods can be viewed as an alternative (though not yet as precise) method that demonstrates the 
need for the existence of states that are not arbitrarily heavy in a two dimensional CFT at large central charge.
Our method relies on consistency of the perturbative $1/c$ expansion rather than unitarity and locality, which are harnessed in the bootstrap.

One important comment is in order, which is that the exact spectrum of states which must be added in the theory according to this argument is rather subtle.  In particular, a naive application of resurgence methods along the lines of what we have discussed turns out to imply that we must add an infinite tower of new primary states with {\it negative} conformal weights:
\be\label{tower}
  h = -m(m-1){c\over 6} + \calo(c^0), \quad m = 2,3,\ldots
\end{equation}
In the case of Virasoro minimal model CFTs with negative central charge, this is indeed exactly what happens -- these are the conformal weights of a universal set of primary states $\phi_{1,2m-1}$ 
that appear in the spectrum of the minimal model CFTs.
For unitary CFTs with positive central charge the interpretation must be different, as the spectrum cannot include states with negative conformal weights.  Instead, in such a CFT states with {\it positive} dimension must be added in such a way as to cancel the divergent asymptotic series.  This will likewise involve states with dimension of order ${\cal O}(c)$, but with an exponentially large density of states and with OPE coefficients which conspire to mock up the effect of the tower of states described in (\ref{tower}).  In such a theory, the tower (\ref{tower}) should be regarded as ``fictitous" states that do not appear in the physical spectrum of the theory, but that nevertheless govern the structure of the perturbative expansion.

This can be made much more explicit in the holographic interpretation of these CFTs at large central charge.
Indeed, in the AdS/CFT correspondence the large $c$ limit is the semi-classical limit of the corresponding theory of gravity.  The semiclassical block $f(x)$ is the (regularized) Einstein-Hilbert action of an appropriate asymptotically AdS geometry, and 
the terms in the $1/c$ expansion describe the loop corrections to this semi-classical result.\footnote{It is important to note that, although general relativity in three dimensions has no local degrees of freedom, the loop expansion is still non-trivial.  In cases where we can compute explicitly, it matches exactly the terms in the $1/c$ expansion of the appropriate conformal block (see e.g. \cite{Giombi:2008vd}).} 
So our CFT result has a simple dual interpretation: the graviton loop expansion is an asymptotic series.  
Although it is generally expected that perturbative expansions in quantum field theory are asymptotic, it is gratifying to see that this is indeed the case in a genuine (albeit simple) theory of gravity.

We can now interpret our result as a statement about the gravitational path integral in three dimensions.  In particular, we see that the graviton loop expansion is an asymptotic series, which implies the existence of new gravitational instantons -- solutions to Einstein's equations.  In a sense, we are using to the perturbative structure of the graviton loop expansion to infer the existence of a non-trivial sum over geometries.
A subtlety arises, however, when we attempt to identify explicitly the geometries which correspond to the tower of states (\ref{tower}). The geometries associated with this tower are conical excesses, singular solutions to the equations of motion that are obtained from empty AdS by changing the periodicity of the angular variable to be
\be\label{excess}
\Delta\phi = 2\pi (2m-1)~.
\ee
Our interpretation is {\it not} that these should be regarded as physical configurations that contribute directly to the path integral of three dimensional gravity, at least not in Euclidean signature. Rather, they should perhaps be interpreted as solutions to the equations of motion that do not lie on the physical contour of integration through the space of metrics.
Nevertheless, their existence controls the asymptotic behaviour of the graviton loop expansion. 
This phenomenon is actually quite familiar from quantum mechanics, where complex instanton solutions can control the behaviour of the perturbative expansion even though they do not describe physical processes.\footnote{A nice analogy, elaborated on below, is to the perturbative expansion of the anharmonic oscillator with potential $V(x) = x^2+\lambda x^4$.  When $\lambda$ is negative, the asymptotic growth of perturbative coefficients is governed by the  solution to the equations of motion which mediates tunnelling. When $\lambda$ is positive we still have an asymptotic growth of perturbative coefficients, but the corresponding solution to the equations of motion is now complex.} 

We should emphasize here that these ``saddles'' do not lie on the contour of integration of the \emph{Euclidean} gravitational path integral, where they would correspond to the exchange of local operators with scaling dimensions that violate the unitarity bound in the OPE. However, the picture in Lorentzian signature is more subtle. In particular, singularities in Lorentzian signature need not correspond to the exchange of local operators. In fact, precisely the tower of conical excess saddles corresponding to (\ref{excess}) was discovered in \cite{Fitzpatrick:2016mjq} by consideration of the monodromy problem for the Virasoro vacuum block. There it was argued that these saddles are not visible on the first sheet but rather are generated upon analytic continuation to higher sheets of the complex cross-ratio plane in Lorentzian signature. Indeed, the existence of these saddles in Lorentzian signature was used to compute the late-time behaviour of the out-of-time-ordered correlator in \cite{Chen:2016cms} by analytic continuation of the Borel-resummation of their contributions to the vacuum block to the second sheet.\footnote{We are grateful to Liam Fitzpatrick for helpful discussions on these points.} In this work we will argue for the existence of the conical excess saddles (\ref{excess}) by demonstrating the existence of the corresponding poles in the Borel plane of the Virasoro vacuum block directly.

Finally, we note that -- given the close relationship between two dimensional conformal field theory and three dimensional Chern-Simons theory -- it is  no surprise that these effects have a clear interpretation in Chern-Simons theory.  Indeed, the fact that the $\frac{1}{k}$ expansion in Chern-Simons theory is an asymptotic series is well-known (see e.g. \cite{Gukov:2016njj}).  This has primarily been studied in the context of compact gauge group on manifolds without boundary, where the saddle points can be identified with flat connections in the usual way.  In our case, Virasoro blocks are related to Chern-Simons theory with gauge group $PSL(2,{\mathbb R})$.  The essential observation, which goes back to \cite{Verlinde:1989ua}, is that Virasoro conformal blocks are obtained by quantizing the moduli space of flat $PSL(2,{\mathbb R})$ connections.  This moduli space has several disconnected components, one of which -- the ``Teichm\"uller component'' -- is related to the moduli space of smooth hyperbolic metrics.  For conformal blocks with sufficiently heavy internal operators, the block can be obtained by quantizing this Teichm\"uller component.  Our result, however, implies that the asymptotic expansion of the vacuum block necessitates the inclusion of non-perturbative effects related to the other components of moduli space.  These non-Teichm\"uller components describe the moduli space of hyperbolic metrics with conical singularities, as anticipated by our 3d gravity result (\ref{excess}).  Roughly speaking, the vacuum block should be regarded not just as a wave function on the Teichm\"uller component, but instead as a wave function on the full moduli space with a (non-perturbatively) small tunnelling amplitude which describes mixing between the Teichm\"uller and non-Teichm\"uller components.

Our asymptotic results hence admit a simple gravitational interpretation: quantum general relativity, defined by quantizing the space of non-singular metrics which are perturbations of the vacuum, cannot be a consistent theory on its own. Its spectrum must include more than perturbative gravitons. This statement is well-understood holographically as a consequence of consistency conditions like unitarity and locality of the dual CFT, but it is satisfying to have derived a clean signature of this pathology in gravitational perturbation theory. 

We emphasize that there are in principle many distinct ways to complete the vacuum block into a well-defined CFT correlator, and correspondingly there may be many UV-complete theories of gravity described at low energies by three-dimensional Einstein gravity. In other words, in the absence of other stringent constraints, the knowledge of the vacuum block alone is wildly insufficient to specify the correlator, let alone the theory. Our analysis shows that any UV completion must resolve the singularities in the Borel plane corresponding to the infinite tower (\ref{tower}), but it is an open problem to specify the precise mechanism by which this is accomplished. Going forward, it would be interesting to understand more concretely how this carves out the spectrum and CFT data of heavy states of holographic CFTs. 

Our paper is organized as follows.  In section \ref{sec:reviewasymptotic} we will review briefly a few features of asymptotic series and Borel summation, and present a simple example which is instructive in what follows.  Section \ref{sec3} contains our main result, on the nature of the asymptotic expansion of the vacuum conformal block.  We will first describe explicitly the computation of conformal blocks in rational CFTs, where everything can be done analytically.  Then we will use Zamolodchikov's recursion relation to generalize these results to genuine non-rational CFTs.  Finally we will summarize the evidence based on the numerical computation of conformal blocks.  Section \ref{sec:grav3d} contains a discussion of the interpretation of these results, first in 3d gravity and then in Chern-Simons theory, along with a discussion of the relationship with the monodromy method for the computation of classical conformal blocks.  We discuss a few open questions in Section \ref{sec:discuss}.  Two appendices contain details of the numerics, along with some details of the interpretation of the non-Teichm\"uller components of the moduli space of $SL(2,{\mathbb R})$ connections.

\section{Review: asymptotic series and the Borel plane}
\label{sec:reviewasymptotic}

We now briefly review of a few aspects of asymptotic series, and how they are used to infer the existence of non-perturbative effects.

Perturbative and non-perturbative effects in physical systems are often intricately related.
In particular, although a perturbation series does not directly include non-perturbative contributions, the asymptotic behaviour of the coefficients in this series can secretly encode the behaviour of non-perturbative effects. 
This is the beautiful idea behind resurgence, which has been applied to study quantum systems in a variety of settings.\footnote{See e.g. \cite{Marino:2012zq,Dorigoni:2014hea, Jentschura:2004jg, Dunne:2014bca} for reviews.} As a simple example, consider a partition function $Z(g)$ computed perturbatively in some coupling constant $g$ 
\begin{equation}
\label{1}
Z(g) \equiv \int D\phi \, e^{-S[\phi]/g} \simeq \sum_{n=0}^{\infty}a_{n} g^{n}
\end{equation}
Here our partition function is a path integral over some degrees of freedom $\phi$ with an action $S[\phi]$.  We have assumed that this theory has a classical minimum with zero action, for simplicity. 
The coefficients $a_n$ in the perturbation series on the right are found by expanding around this zero action solution.
For many systems this is an asymptotic series rather than a convergent one, and at large $n$ the coefficients $a_{n}$ grow factorially with $n$. This rapid growth is usually associated with the growth of the number of Feynman diagrams involved in the computation of $a_n$.  As a result, the series (\ref{1}) has vanishing radius of convergence.

In order to render this series sensible, a typical approach considers instead the Borel series $BZ(S)$ defined as
\begin{equation}
\label{2}
BZ(S) \equiv \sum_{n=0}^{\infty} \frac{a_{n}}{n!} S^{n}
\end{equation}
If $a_n \sim n!$ at large $n$, then this sum has a finite radius of convergence.  In many cases one can then analytically continue $BZ(S)$ to any complex value of $S$ with positive real part, allowing us to think of $BZ(S)$ as a function on the complex $S$ plane (known as the Borel plane).  One can then define the ``Borel resummed" partition function $\tilde{Z}(g)$ by
\begin{equation}
\label{3}
\tilde{Z}(g) = \frac{1}{g} \int_{0}^{\infty} dS \, BZ(S) \, e^{-S/g}
\end{equation}
where we integrate this analytically-continued expression for $BZ(S)$.
Note that when we insert the series expansion for $BZ(S)$ in equation (\ref{2}) into this expression, and interchange the sum and integral, this   reproduces the series expansion for $\tilde{Z}(g)$ in (\ref{1}). This approach is somewhat ambiguous, but 
-- provided we can analytically continue $BZ(S)$ --  
equation (\ref{3}) at least in principle allows us to use the perturbative series to define the partition function at generic values of the coupling constant.  

One problem is that there is no guarantee that $BZ(S)$, once analytically continued, will be a smooth function of the parameter $S$.  For example, $BZ(S)$ may contain a singular point 
at some value $S=S_{0}$ in the Borel plane, 
which by deforming the contour of integration in equation (\ref{3}) would appear to contribute to $\tilde{Z}(g)$. 
For example, if $a_{n}=n!(S_{o})^{-n}$, the Borel transform
\begin{equation}
BZ(S) = \sum_{n=0}^{\infty} \left( \frac{S}{S_{o}}\right)^{n}  = \frac{1}{1-\frac{S}{S_{o}}} = -\frac{S_{o}}{S-S_{o}} \,,\quad (S<S_{o})
\end{equation}
has a simple pole at $S=S_{o}$ in the Borel plane.  Upon deforming the $S$-contour in equation (\ref{3}) this leads to a contribution $e^{-S_o/g}$ to $\tilde{Z}(g)$.  This is the expected form of a non-perturbative contribution to the partition function coming from a classical solution with action $S_o$. 

This is a simple example of a general phenomenon: 
perturbative expansion coefficients typically grow like $a_{n}\sim n!(S_{o})^{-n}$ at large $n$, where $S_o$ is the action of a new saddle point.  We note that this growth need not come from simple pole in Borel plane; any singular point of the type $BZ(S) \sim (S_{o}-S)^{-\alpha}$ will lead to this behaviour, as can be checked by expanding $BZ(S)$ near zero.  Moreover, the behaviour in the Borel plane can be quite complicated and include many poles or branch cuts.  The singular point closest to the origin is the one that will determine the asymptotic behaviour of our $a_n$.
More generally, however,
{\it any} classical saddle point of the action $S[\phi]$ is expected to lead to a singularity in $BZ(S)$.  To see this, we can follow \cite{Lipatov:1976ny, tHooft:1977xjm} to write
\begin{equation}
Z(g) = \int D\phi \, e^{-S[\phi]/g} =  \int D\phi \int_0^\infty dS \, e^{-S/g}\,\delta(S-S[\phi])
\end{equation}
Changing the order of integrations, we can identify the Borel transform as the ``volume in field space" of the surface of fixed action:
\be
BZ(S) = \int D\phi \,  \delta(S-S[\phi])
\ee  
This implies that whenever we have any configuration $\phi_o$ which is a saddle point of the action, we expect a divergence of the Borel transform
\begin{equation}
BZ(S) \sim\left(\frac{\delta S[\phi]}{\delta\phi} \right)^{-1}_{S[\phi_{o}]=S}
\end{equation}
as $S$ approaches $S_o\equiv S[\phi_{o}]$.  

In applying these arguments to general path integrals, one crucial question is whether the singularities in $BZ(S)$ lie on the positive real axis. 
If this is the case, then (\ref{3}) can lead ambiguous results depending on how one deforms the contour of integration to avoid this singularity (see e.g. \cite{Dunne:2016nmc}). In many cases this ambiguity has an important physical interpretation: for example, if we are studying a particle in an unstable potential such terms can be associated to physical effects like false vacuum decay.   For a particle in a stable potential like the double-well, the ambiguous terms need to be removed. The ambiguous terms are typically cancelled by terms coming from other non-perturbative sectors. These cancellations can understood explicitly in some simple quantum mechanical systems \cite{Dunne:2014bca, Lipatov:1976ny, Balitsky:1985in, Dunne:2013ada}.  In these ideal scenarios, where one can compute exactly, one expects that the full result when one includes all non-perturbative effects is well defined and unambiguous.  In more complicated cases, however, one can only constrain the behaviour of non-perturbative effects by studying the behaviour of $BZ(S)$ in the Borel plane, but may not be able to determine $BZ(S)$ unambiguously.

On the other hand, it is also often the case that the singularities in $BZ(S)$ do not lie on the positive real axis.  In this case, there may still be a saddle point $\phi_{o}$ which controls the behaviour of the perturbative expansion.  But it may be a solution (possibly even complex) which does not lie on the physical contour of integration but nevertheless is responsible for determining the behaviour of the asymptotic series.  This will be the case for the perturbative expansion of conformal blocks considered in this paper.  So before proceeding, we now describe in more detail a simple example which illustrates this basic phenomenon.

\subsection{A zero-dimensional example}
We will illustrate these ideas in the case of the following one-dimensional integral:
\begin{equation}\label{integral}
    Z(\hbar) = \int_{-\infty}^{\infty} d\phi \, e^{-S(\phi)/\hbar} ~,\quad S(\phi) = \frac{1}{2}\phi^{2} + \frac{\lambda}{4}\phi^{4}
\end{equation}
This integral is well-defined and finite when $\hbar, \lambda\ge0$.   We can compute $Z(\hbar)$ perturbatively by expanding around the minimum of the action at $\phi=0$.  Expanding the interaction term $e^{\lambda\phi^{4}/4}$ and doing the Gaussian integrals gives
\begin{equation}
    \label{Zpert}
    Z(\hbar) \simeq \sqrt{2} \sum_{n=0}^{\infty} \frac{(-\lambda)^{n}}{n!} \Gamma\left(2n + \frac{1}{2}\right) \hbar^{n+\frac{1}{2}}~.
\end{equation}
So we have an expansion $Z\simeq\sum_{n}a_{n}\hbar^{n+\frac{1}{2}}$ with perturbative coefficients
\begin{equation}\label{ais}
    a_{n} = \sqrt{2}\, \frac{(-\lambda)^{n}\Gamma\left(2n + \frac{1}{2}\right)}{n!} \sim (-4g)^{n}\, n! \quad (n\gg 1)
\end{equation}
that grow factorially, as expected. 
This is the analog of the Feynman diagram expansion of our simple integral, and the coefficient $a_n$ just counts the number of ways to pairwise connect $4n$ points.

Given (\ref{Zpert}), we define the Borel sum
\begin{equation}
    \label{BZ}
    BZ(S) = \sqrt{2} \sum_{n=0}^{\infty} \frac{(-\lambda)^{n}}{n! \Gamma(n+3/2)} \Gamma\left(2n + \frac{1}{2}\right) S^{n+1/2}
\end{equation}
so that
\begin{equation}
    Z(\hbar) = \frac{1}{\hbar} \int_{0}^{\infty} dS \, e^{-S/\hbar} \, BZ(S)~.
\end{equation}
We recover the perturbative expansion by substituting (\ref{BZ}) and exchanging the sum and the integral. 
In this case the series (\ref{BZ}) can be summed explicitly, and we find
\begin{equation}
    BZ(S) = 2\sqrt{\frac{\sqrt{1+4\lambda S}-1}{\lambda}} = 2 \left. \frac{dS(\phi)}{d\phi}\right|_{S=S(\phi)}^{-1}~.
\end{equation}
We see that that there are singular points (branch cuts) in the Borel plane at $S=-1/4\lambda$. 
As anticipated, these correspond to classical solutions of the equations of motion.
Indeed, the equation of motion
\begin{equation}
    S'(\phi_{o}) = \phi_{o}\left(1+ \lambda\phi_{o}^{2}\right) = 0 
\end{equation}
has two complex solutions at $\phi_{o} = \pm \frac{i}{\sqrt{\lambda}}$ in addition to the original $\phi_o=0$ solution we expanded around.  The action of these solutions is $S(\phi_{o}) = -1/4\lambda$.  Indeed, we see that this classical action was visible in our original expansion (\ref{ais}), since $a_n \sim (-4\lambda)^{n} n! = S(\phi_o)^{-n} n!$.

This simple example illustrates that the exponential piece of the perturbative coefficients encodes the classical action of a new saddle point.
It also illustrates another important point: the saddle points which arose in this way were complex, and did not lie on the original contour of integration of (\ref{integral}).  Thus complex saddle points can control the behaviour of perturbation theory even if they do not lie on the physical contour.
Note, however, that if we were to take $\lambda<0$ then these saddles would indeed lie on the contour of integration.  In this case the potential would be unstable, and these solutions would be the zero-dimensional analog of instantons which mediate the decay of the false vacuum at $\phi=0$.  This simple example mirrors closely what will occur for the vacuum conformal block in CFT$_2$: we will discover an asymptotic series which is controlled by ``instanton" effects that correspond to primary operators with negative dimension.  When the central charge is negative, these are genuine primary operators in any (non-unitary) family of minimal models with $c\to-\infty$.  But for positive central charge, they are the CFT analog of the ``complex" instantons which lie off the physical contour of integration, but nevertheless control the perturbative expansion at large $c$.\footnote{A second, less trivial analogy would be to the computation of the ground state energy of the anharmonic oscillator with potential $V(x) = x^2+ \lambda x^4$ in time independent perturbation theory.  When $\lambda>0$ there are no tunnelling effects, but the series is nevertheless asymptotic.  The growth of the coefficients is governed by a complex solution to the anharmonic oscillator equations of motion.  If we were to take $\lambda<0$, this solution becomes the usual physical tunnelling solution that describes the decay of the ground state.}

\section{Resurgence for conformal blocks in CFT$_2$}
\label{sec3}

We will now turn to a study of the asymptotic expansion of Virasoro conformal blocks in powers of $1/c$.
Our primary objects of interest will be the conformal blocks that appear in the computation of sphere $4$-point functions. We will consider the $4$-point function of primary operators $O_{i}(x)$ with conformal dimension $(h_{i},\bar{h}_{i})$ inserted at points $x_i$. 
Using global $SL(2,\mathbb{C})$ transformations to set three of the operator locations to be at $0, 1$ and $\infty$, this
four point function has a conformal block expansion
\begin{equation}
\label{4ptblock}
\begin{split}
\langle O_{1}(\infty)O_{2}(1)O_{3}(z)O_{4}(0) \rangle &= \lim_{w\rightarrow\infty} w^{2h_{1}}\bar{w}^{2\bar{h}_{1}} \langle O_{1}(w)O_{2}(1)O_{3}(z)O_{4}(0) \rangle \\
&= \sum_{p} \frac{C_{12p}C_{34p}}{z^{h_{3}+h_{4}}\bar{z}^{\bar{h}_{3}+\bar{h}_{4}}} \, F(c,h_{i},h_{p},z) \bar{F}(c,\bar{h}_{i},\bar{h}_{p},\bar{z})  \\
\end{split}
\end{equation}
where $z=\frac{x_{12}x_{34}}{x_{13}x_{24}}$ is the cross ratio of the original four points.
The $C_{ijk}$ are OPE coefficients, and the $F(c,h_{i},h_{p},z)$ are Virasoro conformal blocks which package together the contribution of 
an intermediate primary operator $O_{h_{p},\bar{h}_{p}}$ and its descendants. 
Each individual conformal block can itself be written as an expansion in the cross-ratio 
\begin{equation}
\label{9}
F(c,h_{i},h_{p},z) = z^{h_{p}}\sum_{k=0}^{\infty} F_{k}(c,h_{i},h_{p}) \, z^{k}
\end{equation}
Here $F_k$ is the contribution from the descendants at level $k$, and is completely determined by the Virasoro algebra.  
In the large central charge limit conformal blocks can be expanded as\footnote{This form of the semiclassical block assumes that the external and internal operators have conformal weights of order $c$ in the large-$c$ limit.}
\begin{equation}
	\label{10}
	F(c,h_{i},h_{p},z) \simeq e^{cS^{(p)}}\sum_{n\ge0} \frac{a_{n}^{(p)}}{c^{n}}
\end{equation}
where $S^{(p)}$ is the so-called ``classical" conformal block. 
That the leading term exponentiates was first conjectured  in \cite{Zamolodchikov1986} (and later proven in \cite{Besken:2019jyw}), and the fact that the subleading terms take the form of a power series in $1/c$ can be seen by inspecting the individual coefficients $F_k$.
The $4$ point functions then take the form\footnote{In writing the four-point function entirely in terms of semiclassical blocks in this way we have implicitly assumed that the lightest operator that appears in the OPE has twist of order $c$.}
\begin{equation}
\label{11}
\langle O_{1}(\infty)O_{2}(1)O_{3}(z)O_{4}(0) \rangle \simeq \sum_p \frac{C_{12p}C_{34p}}{z^{h_{3}+h_{4}}\bar{z}^{\bar{h}_{3}+\bar{h}_{4}}} \,  e^{cS^{(p)}+\bar{c}\bar{S}^{(p)}}\left( \sum_{n\ge0} \frac{a_{n}^{(p_{1})}}{c^{n}} \right)\left( \sum_{n\ge0} \frac{\bar{a}_{n}^{(p_{1})}}{{c}^{n}} \right) 
\end{equation}
This may be regarded as a sort of ``trans-series'' expansion, where we include not only the infinite perturbative series coming from the expansion around the leading saddle (at sufficiently small cross-ratio, this is the semiclassical conformal block associated with the lightest operator in the OPE) but also the infinite series of non-perturbative corrections associated with the exchange of higher-dimension operators in the OPE.
At sufficiently small cross-ratio the leading contribution will be from the identity conformal block, at least if we take $h_1=h_2$ and $h_3=h_4$. This then motivates the question: can the other primary states $p$, interpreted now as ``subleading saddles," be detected by investigating the perturbative expansion around the leading saddle (i.e. the vacuum state)?  
We will argue that the answer to this question is yes.

In order for this proposal to work, equation (\ref{10}) must be an asymptotic series in $1/c$.  In particular, the coefficients $a_n^{(0)}$ of the vacuum block must grow as 
\begin{equation}
\abs{a_{n}^{(0)}}\sim n! S_{o}^{-n} 
~~~~(n\gg1)
\label{eq:anlargestuff}
\end{equation}
where $S_o$ is the action of one of the subleading saddle points, i.e. it is the classical action associated with the semiclassical block of a {\it different}, non-identity primary state.
Rearranging this equation, we see that at large $n$ the ratio $\frac{1}{n}\ln \frac{\abs{a_{n}^{(0)}}}{n!}$ should approach a constant $S_o$ which is the location of a singularity in the Borel plane. It is important to note that both the coefficients $a_{n}$ and the classical block $S_o$ in this formula are functions of $z$.  In practice, however, we will take $z$ to be small so that we can work at a finite order in the cross-ratio expansion. 
In this limit we expect the singularity in the Borel plane to correspond to the 
contribution from the exchange of another primary in the four-point function, so 
\begin{equation}
\label{13}
e^{-cS_{0}} \sim z^{h_{p}} 
\implies \frac{h_{p}}{c} \sim -\frac{S_{o}}{\ln z}
\end{equation}
This proposal implies that, by understanding the behaviour of the coefficients $a_{n}^{(0)}$ at large $n$, we can infer the existence of other primary states in the theory.  In particular, combining equations (\ref{eq:anlargestuff}) and (\ref{13})
we see that our new primary state has dimension
\be
h_p \sim -\frac{c}{\ln z} \lim_{n\to\infty} \exp\left\{ -\frac{1}{n}\ln \frac{\abs{a_{n}^{(0)}}}{n!} \right\}
\ee
That the expansion coefficients $a_n^{(0)}$ take this form -- with the stated scaling at large $n$ and small $z$ -- is quite nontrivial.  But we will see that this is indeed the case.  
We will focus on the case $h_{1}=h_{2}$ and $h_{3}=h_{4}$, so that the first intermediate primary is the universal contribution coming from the identity operator. In this case (\ref{13}) is a prediction based on exchange of just the vacuum.  

We now carry out this analysis explicitly.  
We will first work with theories with large {\it negative} central charge, where this can be evaluated explicitly in some simple examples.  We will see that the analysis proceed exactly as described above.  We will then turn to a more general analysis using Zamolodchikov recursion relations.  We will see that the analysis leads, in a certain approximation, to an infinite set of poles in the Borel plane.  However, the implications are somewhat more subtle -- the resulting primary states should be regarded as contributions which do not lie on the physical integration contour of the theory, but nevertheless control the structure of the perturbative expansion.
Finally, we summarize the numerical evidence for this picture (presented in more detail in appendix \ref{sec:numerics}).

\subsection{Warm-up: a tractable example}
Here, we consider a 4-point function of degenerate fields $\phi_{r,s}$ in the Virasoro algebra. What we will show in this subsection is that resurgence techniques will automatically ``discover" new operators in the theory.  Indeed, the operators which are discovered in this way are exactly 
the ones that show up in the operator product expansion of $\phi_{r,s} \times \phi_{r,s}$. For concreteness, we will focus on the degenerate operator $\phi_{2,1}$ which has OPE 
\begin{equation}
\phi_{2,1} \times \phi_{2,1} = 
\id+ \phi_{3,1},
\end{equation}
and we will ``discover" the $\phi_{3,1}$ operator from the asymptotic expansion of the vacuum block.  

To begin, the Coulomb-gas formalism allows one to write conformal blocks as contour integrals. For $\phi_{2,1}$ we have
\begin{equation}
\label{phi21}
\begin{split}
\langle \phi_{2,1}(0) \phi_{2,1}(z) \phi_{2,1}(1) \phi_{2,1}(\infty) \rangle &=  \sum_{i=1,2} \frac{C_{i}^{2}}{z^{2h_{2,1}}\bar{z}^{2\bar{h}_{2,1}}} \, F(h_{i},z)\bar{F}(\bar{h}_{i},\bar{z}) \\
\end{split}
\end{equation}
where
\begin{equation}\label{eq:define C and hrs}
c = 13 + 6(b^{2} + b^{-2}) \,,\quad 
h_{r,s} = -{1\over 4b^2}(s^2-1) -{rs-1\over 2} - {b^2\over 4}(r^2 -1).
\end{equation}
In particular we have $h_{2,1} = -{3b^2+2\over 4}$.
The $C_i^2$ are the non-negative squared structure constants describing the fusion $\phi_{2,1}\times \phi_{2,1}$ with the internal operators and $F(h_{i},z)$ are the corresponding conformal blocks. The two intermediate primaries appearing in (\ref{phi21}) are $\phi_{1,1}$ (i.e. $\id$) and $\phi_{3,1}$.  In writing $F(h_{i},z)$, we suppress the dependence on $h_{2,1}$ since it is fixed from now on. In this example either the dimension $h_{2,1}$ is negative or the central charge is negative.  So we are for now considering a non-unitary theory.  
In particular we will be interested in the limit of large (negative) central charge.

The vacuum block admits an integral expression as
\begin{equation}
\label{16}
\begin{split}
	F(h_{1,1}=0,z) &= (1-z)^{-b^{2}/2} \int_{0}^{1} dx \, [x(1-x)(1-zx)]^{b^{2}} \\
\end{split}
\end{equation}
The large $c$ limit can be found by taking $b^{2}$ large, with $c\sim 6b^{2}$. So, we can study the integral in this limit and obtain a series expansion in $1/b^{2}$ or $1/C$ with $C\equiv b^{2}$. Let us focus on the integral without the prefactor.
\begin{equation}
I = \int_{0}^{1} dx \, [x(1-x)(1-zx)]^{b^{2}}
= \int_{0}^{1} dx\, e^{C \psi(x) }
\end{equation}
where $\psi(x) = \ln(x(1-x)(1-zx))$. The critical points of $\psi(x)$ are 
\begin{equation}
\psi'(x)= \frac{1}{x} - \frac{1}{1-x} - \frac{z}{1-zx} =0 \implies x_{\pm} = \frac{1+z \pm \sqrt{z^{2}-z+1}}{3z}
\end{equation}
These saddle points will contribute to $I$ as $\sim  \exp(C \psi(x_{\pm}))$. For $0<z<1$ only $x_{-}$ lies in the range of the integral. We want to evaluate the integral in the large $C$ limit. Extracting the leading piece, we define
\begin{equation}
I' \coloneqq  e^{-C \psi(x_{-})} I  =  \int_{0}^{1} dx\, e^{C(\psi(x) - \psi(x_{-})) }
\end{equation}
To connect with the previous discussion regarding Borel summation, we anticipate that we can write the integral $I'$ in terms of its Borel transform
\begin{equation}
  I' = \int_0^\infty dS \, e^{-CS}BZ(S)
\end{equation}
provided that the integral is well-defined, in other words in the absence of singularities of $BZ(S)$ on the positive $S$ axis.
For a one-dimensional integral this is completely straightforward, and we can explore the analytic structure of the inferred Borel sum $BZ(S)$.  
We perform a change of variables 
\begin{equation}
S(x) = -( \psi(x)-\psi(x_{-}) ) \implies e^{-S} = \frac{x(1-x)(1-zx)}{x_{-}(1-x_{-})(1-zx_{-})}
\end{equation}
For $0<z<1$, $0<x_{-}<1$ so we can split the integral $I'$ as
\begin{equation}
\begin{split}
I' &= \int_{0}^{x_{-}} dx\, e^{-CS(x) } + \int_{x_{-}}^{1} dx\, e^{-CS(x)) } \\
&= \int_{\infty}^{0} dS\, e^{-CS } \, \left[\frac{dS}{dx}\right]^{-1}_{x=x_{1}(S)}  + \int_{0}^{\infty} dS \, e^{-CS } \, \left[\frac{dS}{dx}\right]^{-1}_{x=x_{2}(S)} \\
\end{split}
\end{equation}
where we have two different functions $x_{1}(S)$ and $x_{2}(S)$ appearing since the map from $x\rightarrow{S(x)}$ is not one to one and we need to choose the appropriate solution in the region of integration.  Then
\begin{equation}
\begin{split}
I' &= \int_{0}^{\infty} dS \, e^{-CS } \,\left( - \left[\frac{dS}{dx}\right]^{-1}_{x=x_{1}(S)} + \left[\frac{dS}{dx}\right]^{-1}_{x=x_{2}(S)} \right) \\
&= \int_{0}^{\infty} dS \, e^{-CS } \, BZ(S) \\
\end{split}
\end{equation} 
where we have introduced
\begin{equation}
\label{24}
BZ(S) = - \left[\frac{dS}{dx}\right]^{-1}_{x=x_{1}(S)} + \left[\frac{dS}{dx}\right]^{-1}_{x=x_{2}(S)}
\end{equation}
Expanding $BZ(S)$ around $S=0$
\begin{equation}
BZ(S) = \sum_{n=0}^\infty c_{n}S^{n-1/2}
\end{equation}
is the desired Borel transform. The above sum when inserted in the previous integral gives the asymptotic series expansion for $I'$:
\begin{equation}
I' 
= \int_{0}^{\infty} dS \, e^{-CS } \sum_{n=0}^{\infty} c_{n} S^{n-1/2} \\
=  \sum_{n=0}^{\infty} \frac{c_{n}\Gamma(n+1/2)}{C^{n+1/2}} \\
\end{equation}
Note that $I'$ has an expansion in powers of $1/C^{n+1/2}$ rather than $1/C^{n}$ because the leading term comes from a Gaussian integral, giving a factor of $1/\sqrt{C}$.
Putting this together, we find
\begin{equation}
I = e^{C \psi(x_{-})} \int dx e^{C \frac{\psi''(x_{-})}{2} (x-x_{-})^{2}} + \dots =  \frac{e^{C \psi(x_{-})}}{\sqrt{-C \frac{\psi''(x_{-})}{2}}}\sum_{n=0}^\infty \frac{a_{n}}{C^{n}}
\end{equation}
where the coefficients $a_{n}$ are proportional to $c_{n}\Gamma(n+1/2)$.  This makes manifest the factorial growth of the coefficients.

We can now compute the coefficients exactly and extract the power law behaviour $S_{o}^{-n}$.
Defining
\begin{equation}
\label{28}
x_{1}(S) = \frac{1+z}{3z} - \frac{\sqrt{1-z+z^{2}}}{3z}\, 2\cos\frac{\theta}{3}
\end{equation}
\begin{equation}
\label{29}
x_{2}(S) = \frac{1+z}{3z} + \frac{\sqrt{1-z+z^{2}}}{3z}\,  \left( \cos\frac{\theta}{3} - \sqrt{3}\sin\frac{\theta}{3} \right)
\end{equation}
where
\begin{equation}
\begin{split}
b &= \abs{b} e^{i\theta} = a + i \sqrt{4z^{12}(1-z+z^{2})^{3} - a^{2} } \\
a &= z^{6} \left[ -(z+1)(z-2)(2z-1) + e^{-S}\{ (z+1)(z-2)(2z-1) - 2 (1-z+z^{2})^{3/2}  \} \right]  \\
\end{split}
\end{equation}
(\ref{24}) can now be expanded near $S=0$ to arbitrary order. 
It is straightforward to verify that the result correctly reproduces the value
\begin{equation}
S_{o} = -( \psi(x_{+})-\psi(x_{-}) )
\end{equation}
of the other saddle point action for the integral $I'$.  We note that in this expression $S_{o}$ is still a function of $z$; in fact, it is the classical conformal block for the operator $\phi_{3,1}$. 

To see this a bit more simply, note that the saddles $x_{\pm}$ have physical interpretations in terms of the primary operators $\phi_{1,1}$ and $\phi_{3,1}$.  Specifically, as $z\rightarrow 0$ we have
\begin{equation}
x_{-} \sim \frac{1}{2}\,, \quad x_{+} \sim \frac{2}{3z}
\end{equation}
which means the conformal blocks (\ref{16}) scale as
\begin{equation}
F_{-} \sim 1 \,,\quad F_{+} \sim z^{-2b^{2}}
\end{equation}
at small $z$, where the $\pm$ subscript denotes that we are evaluating at $x_{\pm}$. Comparing the above with the expected scaling of conformal blocks in four-point functions (\ref{9}) we find that the $x_{-}$ corresponds to the vacuum block (the intermediate primary with $h=0$) whereas $x_{+}$ corresponds to a primary with $h=-2b^{2}$.  In the large $b$ limit this matches the dimension of the operator $h_{3,1}=-2b^{2}-1$. So the resurgence analysis has reproduced the fusion rule 
\begin{equation}
\label{34}
\phi_{2,1} \times \phi_{2,1} = 
\id+ \phi_{3,1},
\end{equation}
where $\phi_{1,1}$ is the vacuum primary. To summarize: we have computed the large $b$ limit of the vacuum block (since we were computing $1/C$ perturbation theory around the $x_{-}$ saddle) and found that the perturbative coefficients diverged factorially in exactly the way needed to reproduce the subleading classical block $e^{-b^{2}S_{o}}$.

This example makes clear that the large $c$ expansion of the vacuum conformal block contains information about the spectrum and OPE coefficients of non-vacuum primaries. We now would like to carry out the above analysis for general blocks where we do not have an exact integral representation using the Coulomb gas formalism. 

\subsection{Generic 4-point blocks using recursion relations}

To carry out the more general analysis, we will use Zamolodchikov's recursion relations \cite{1987TMP....73.1088Z, 1984CMaPh..96..419Z}. These provide an efficient way to compute the coefficients $F_{k}(c,h_{i},h_{p})$ in (\ref{9}). There are two different ways of writing these recursion relations.  We will use the so-called ``$c$-recursion" relation: 
\begin{equation}
\label{36}
F(c,h_{i},h,z) = z^{h} \, {}_{2}F_{1}(h-h_{12},h+h_{34},2h;z) \,  +  \sum_{m\ge1,n\ge2}^{\infty} \frac{R_{mn}(h_{i},h)}{c-c_{mn}(h)}F(c_{mn}(h),h_{i},h+mn,z)
\end{equation}
where $h_{ij}=h_{i}-h_{j}$ and 
\begin{equation}
c_{mn}(h) = 13 - 6( t_{mn} + t_{mn}^{-1} )
\end{equation}
\begin{equation}
t_{mn} = \frac{2h + mn - 1 + \sqrt{ 4h(h + mn - 1) + (m-n)^{2} }  }{n^{2}-1}
\end{equation}
\begin{equation*}
    \begin{split}
    &R_{mn}(h_{i},h) = D(h) \frac{\prod_{j,k}\left(l_{2}+l_{1}-l_{jk}/2 \right) \left(l_{2}-l_{1}-l_{jk}/2 \right)\left(l_{3}+l_{4}-l_{jk}/2 \right)\left(l_{3}-l_{4}-l_{jk}/2 \right)}{\prod_{a,b}\, l_{ab}}\\
    &j = -(m-1), -(m-3), \dots, m-3, m-1 \,,\quad k = -(n-1), -(n-3), \dots, n-3, n-1 \\
    &a = -(m-1), -(m-2), \dots, m-2, m-1, m \,, \quad b = -(n-1), -(n-2), \dots, n-2, n-1, n \\
    \end{split}
\end{equation*}
\begin{equation*}
    D(h) = \frac{12( t_{mn} - t_{mn}^{-1} )}{(m^{2}-1)t_{mn}^{-1} - (n^{2}-1)t_{mn}} \,, \quad l_{i} = \sqrt{h_{i} + l_{11}^{2}/4 } \,, \quad l_{jk}  = \frac{j-k \, t_{mn}}{\sqrt{t_{mn}}}
\end{equation*}
In this product, the pair $(a,b)=(0,0),(m,n)$ are omitted. The recursion relation (\ref{36}) was obtained by Zamolodchikov \cite{1987TMP....73.1088Z, 1984CMaPh..96..419Z} by considering the conformal blocks as analytic function of the central charge $c$ and observing that the coefficients have simple poles when $c$ takes a value $c_{mn}$ where the Virasoro algebra has a null state. The first term on the RHS in (\ref{36}) is the seed of the recursion and is the global conformal block that one recovers in the $c\rightarrow\infty$ limit with dimensions of external operators fixed. 
The terms appearing in the sum on the RHS of (\ref{36}) are proportional to $z^{h+mn}(1 + \mathcal{O}(z)\, )$, so to compute the coefficients on the LHS up to a finite order in $z$ we only need finite number of terms on the RHS. 

We are interested in the vacuum block, so will take our 
external operators to be pairwise equal. Using the recursion relation (\ref{36}), we can then obtain an asymptotic series in $1/c$ and study the growth of the perturbative coefficients. 
Performing an analysis as in the previous example, the result should -- at least in principle -- indicate a new primary $\phi_{h}$ that will contribute to the four-point function.  In particular, we expect 
\begin{equation}
\label{39}
h = d c \quad (\mbox{for large } c)
\end{equation}
where $d$ is constant extracted from the behaviour of the $a_{n}$ at large $n$ but small $z$.

\subsubsection*{Asymptotic series from recursion relations}
The general structure of the (vacuum) conformal block is of the form 
\begin{equation}
    F_{vac}(c,z) = \sum_{m\ge1,n\ge2} \frac{R_{mn}F_{mn}(z)}{c-c_{mn}} =  \sum_{m\ge1,n\ge2} \frac{R_{mn} \, z^{mn}(1+\dots)}{c-c_{mn}}
    \label{eq:fvacqenfirst}
\end{equation}
The crucial point is that the only dependence on $c$ is in the denominators $c-c_{mn}$ for fixed $h_{i}$.  For the vacuum block we have
\begin{equation}
    c_{mn} = 13 - 6\left(\frac{m+1}{n+1} + \frac{n+1}{m+1}\right)
\end{equation}
For instance $c_{12}=0$ and $c_{14}=-22/5$. We also have \cite{Perlmutter:2015iya}
\begin{equation}
    R_{mn}=0 \,, \quad \mbox{for $mn$  odd or $m\ge n$}  
\end{equation}
so we have (note that we shifted $m, n$ by $1$ going from (\ref{eq:fvacqenfirst}) to (\ref{eq:fvaceqnsec})):
\begin{equation}
    \begin{split}
    F_{vac}(c,z) &\approx \sum_{n\ge 3}\sum_{\,2\le m\le n} \frac{R_{m-1,n-1}\, z^{(m-1)(n-1)}}{c-13+6\left(\frac{m}{n} + \frac{n}{m}\right)} \\
    &= \frac{1}{c}\sum_{k\ge0}\frac{1}{c^{k}} \sum_{n\ge 3}\sum_{\,2\le m\le n} R_{m-1,n-1}\, z^{(m-1)(n-1)}\left(13 - 6\left(\frac{m}{n} + \frac{n}{m}\right) \right)^{k}. \\
    \end{split}
    \label{eq:fvaceqnsec}
\end{equation}
Note that in going from  (\ref{eq:fvacqenfirst}) to (\ref{eq:fvaceqnsec}) we have approximated the function $F_{mn}(z)$ as just $z^{mn}$. The corrections (labeled as $\ldots$ in (\ref{eq:fvacqenfirst})) are subleading at large $k$ in the asymptotic $c^{-k}$ expansion as long as $z$ is small. In particular, the coefficients $a_{k}$ of $c^{-k}$ are
\begin{equation}
    \begin{split}
        a_{k} &\approx \sum_{n\ge 3}\sum_{\,2\le m\le n} \exp[ \log(R_{m-1,n-1}) + (m-1)(n-1)\log(z) + k \log(13 - 6\left(\frac{m}{n} + \frac{n}{m}\right))  ] \\
        &\approx \sum_{n\ge 3} \exp[ \log(R_{1,n-1}) + (n-1)\log(z) + k \log(13 - 6\left(\frac{2}{n} + \frac{n}{2}\right))  ]
    \end{split}
\end{equation}
where in going to the second line we set $m=2$; note that since $z\rightarrow0$ the larger values of $m$ are suppressed. Performing a saddle point analysis to approximate the sum
\begin{equation}
    \log(z) + k\frac{\frac{12}{n^{2}} - 3 }{13 - 6\left(\frac{2}{n} + \frac{n}{2}\right)} + \frac{1}{R_{1,n-1}} \frac{d R_{1,n-1}}{dn}  = 0
\end{equation}
at large $k$ and $z\rightarrow0$ only the first two terms contribute
\begin{equation}
    \log(z) + \frac{k}{n} \approx 0 \implies n = \frac{k}{-\log(z)}
\end{equation}
where we have also taken $n$ large.  This holds as long as $k \gg -\log(z)$. Plugging this value of $n$ in the exponent, we get:
\begin{equation}
    a_{k} \sim \exp[-k + k \log(\frac{3k}{\log(z)}) ] = \left( \frac{k}{e}\right)^{k}  \left( \frac{\log(z)}{3} \right)^{-k} \sim k! \left( \frac{\log(z)}{3} \right)^{-k}
\end{equation}
We conclude that the $1/c$ expansion is an asymptotic series due to the factorial growth in the coefficients. Moreover, we see that $a_{k} \sim k! S_{o}^{-k}$ where
\begin{equation}
    e^{-cS_{0}} = e^{-\frac{c\log(z)}{3}} = z^{-\frac{c}{3}}
\end{equation}
This is the leading term in the small $z$ expansion of a conformal block with dimension of the intermediate primary given to leading order by $-c/3$. In fact, this is the dimension of $h_{3,1}$ at large $c$, just as in the previous section!  So our generic analysis using the recursion relations reproduces the specific result derived using the Coulomb gas integrals. In fact, we will now show that keeping the other terms in sum over $m$ one can also obtain the other $h_{r,1}$ (for $r$ odd) operators.

\subsubsection*{$h_{r,1}$ from recursion relations}

In the above we included the term $m=2$ and found the saddle point in the sum over $n$. We now generalize this to any $m$. To find the saddle point in $n$, we need
\begin{equation}
    (m-1)\log(z) + k\frac{\frac{6m}{n^{2}} - \frac{6}{m} }{13 - 6\left(\frac{m}{n} + \frac{n}{m}\right)} + \frac{1}{R_{m-1,n-1}} \frac{d R_{m-1,n-1}}{dn}  = 0
\end{equation}
Again we consider large $k$ and small $z$
\begin{equation}
    (m-1)\log(z) + \frac{k}{n} = 0 \implies \frac{k}{-(m-1)\log(z)}
\end{equation}
Substituting this in $a_{k}$
\begin{equation}
    a_{k} \sim \exp( -k + k \log(k) + k\log(\frac{6k}{m(m-1)\log(z)})  ) \sim k! \left( \frac{m(m-1)\log(z)}{6} \right)^{-k}
\end{equation}
which gives the saddle point action as $S_{o} = \frac{m(m-1)\log(z)}{6}$. The dimension of the exchanged primary is 
\begin{equation}
    h = -m(m-1)\frac{c}{6}
    \label{eq:generalmhstuff}
\end{equation}
These are precisely the dimension of operators $h_{r,1}$ (for odd $r = 2m-1$) at large $c$:
\begin{equation}
    h_{r,1} = (1-r^{2})\frac{c}{24} + \mathcal{O}(1)
    \label{eq:hr1stuff}
\end{equation}
From the point of view of AdS$_{3}$ such operators correspond to geometries with a conical excess. This will be discussed in more detail in section \ref{sec:grav3d}. 
The connection between the conical excess geometries and such degenerate conformal weights has been noticed before \cite{Campoleoni:2013lma,Raeymaekers:2014kea,Raeymaekers:2020gtz}. Indeed in \cite{Raeymaekers:2014kea,Raeymaekers:2020gtz} fluctuations around the conical excess geometries were studied in 3d gravity, and it was shown that quantization leads to the same nonunitary representations of the Virasoro algebra, including the correct one-loop correction to (\ref{eq:hr1stuff}) implied by (\ref{eq:define C and hrs}).

Note that unlike in the previously discussed toy example of non-unitary minimal models, where the operator $\phi_{3,1}$ explicitly shows up in the OPE of $\phi_{2,1}$ with itself (see e.g. (\ref{34})), we again emphasize that we are \emph{not} implying that operators of dimensions in (\ref{eq:hr1stuff}) must show up in the partition function of a large $c$ 2d CFT, or in the gravitational path integral. Indeed these operators have negative dimension so their existence would be in contradiction with unitarity of the CFT. Instead, we argue terms in (\ref{eq:hr1stuff}) are analogous to solutions of the equations of motion off of the physical contour of integration. The true spectrum of a theory at large $c$ will reproduce the effect of these operators even though they do not show up in the physical spectrum of the theory.  This precisely mirrors the behaviour of the anharmonic operator presented at the end of section \ref{sec:reviewasymptotic}: when $c>0$, the operators $h_{r,1}$ are the analog of the complex saddle points which are found by analytic continuation of the tunnelling solution from negative $\lambda$ to positive $\lambda$.

\subsection{Numerics}

In appendix \ref{sec:numerics}, we use the recursion relations to test this resurgence numerically. We use the code of \cite{Chen:2017yze} which allows us to compute conformal blocks for pairwise identical external operators to very high precision.  We look explicitly at the vacuum block of four identical external operators in three cases: four degenerate operators $\phi_{2,1}$, four light operators $\phi_{h_L}$ whose dimension do not scale with $c$, and four heavy operators $\phi_{h_H}$ whose dimension scales with $c$. 

Our basic strategy is the following. Using the code in \cite{Chen:2017yze}, we compute the vacuum conformal block to high order in powers of $q$ (a function of the cross-ratio $z$, see appendix \ref{sec:numerics} for more details), for specific external operators at arbitrary central charge $c$. We numerically evaluate this at a fixed $q$ so that our (truncated) $h$-recursion gives a reliable numerical result and then look at the vacuum block explicitly in $1/{c^n}$ expansion. The expansion is asymptotic so we perform a Borel transform numerically, and determine the radius of convergence by observing the growth of the Borel-transformed asymptotic expansion at large $n$, as in Eqns (\ref{eq:anlargestuff}), (\ref{13}). For more details see appendix \ref{sec:numerics}.

In the case of four degenerate operators, the Borel transform is known exactly. However we also use the recursion relations as check of our numerics, which will be useful because for generic external operators we do not have exact expressions. For both four light external operators and four heavy external operators, we show that the numerics are consistent with an operator exchanged with dimension $-c/3$. This is the first of the operators predicted in (\ref{eq:generalmhstuff}) (with $m=2$). Note that because the numerics are rather delicate (involving truncation in the $h$-recursion and numerically fitting and extrapolating the Borel-transformed series to large $n$), we can only match the expected result $-c/3$ within a few percent. It would be interesting to refine these methods to systematically improve this, and recover the full classical block as well as the subleading poles.

\section{3D gravity and $PSL(2,\mb{R})$ Chern-Simons theory interpretation}
\label{sec:grav3d}

A two dimensional CFT can naturally be interpreted as a theory of gravity in three dimensional anti-de Sitter space, with the central charge playing the role of the coupling constant $c=\frac{3 \ell_{AdS}}{2G_N}$ that controls the strength of gravitational interactions.  
Indeed, the classical block which dominates at large $c$ is precisely the (regularized) Einstein action of an appropriate bulk solution (see for example \cite{Hartman:2013mia,Faulkner:2013yia}).
The $1/c$ expansion naturally plays the role of a graviton loop expansion, and our asymptotic analysis implies that new saddle points will arise in the path integral of AdS$_3$ gravity.  We will discuss these saddle points in the next section.  We will also discuss the interpretation in the language of $PSL(2,{\mathbb R})$ Chern-Simons theory.

\subsection{Non-perturbative effects in Virasoro conformal blocks and conical excesses in 3D gravity}

In the previous section 
we found evidence from the asymptotic $1/c$ expansion of the sphere four-point vacuum block of non-perturbative effects associated to the exchange of primaries with conformal weights
\begin{equation}\label{eq:excessWeight}
  h = -m(m-1){c\over 6} + \calo(c^0), \quad m = 2,3,\ldots
\end{equation}
In the semiclassical limit, such operators correspond to massive point particles that backreact on the geometry and source geometries with a conical singularity in AdS$_3$ gravity. That the conformal weight (\ref{eq:excessWeight}) is negative reflects the fact that the non-perturbative effects we have discovered actually correspond to conical excesses rather than conical defects. The conical excess angle is given in terms of the conformal weight as\footnote{In what follows we will focus on the case of massive scalar excess operators, so that $\bar h = h$.}
\begin{equation}
    \frac{\Delta \phi}{2\pi} = \sqrt{1 - \frac{24h}{c}}
\end{equation}
where $\phi \sim \phi + \Delta\phi$. Considering the case of excess angles quantized in integer multiples of $2\pi$, $\Delta\phi_n = 2\pi n$, we have
\begin{equation}
   h_n \coloneqq (1-n^{2})\frac{c}{24}.
\end{equation}
So we see that the dimensions of operators that source conical excess angles that are an odd multiple of $2\pi$
\begin{equation}
    h_{2m-1} = (1 - (2m-1)^{2})\frac{c}{24} = -m(m-1)\frac{c}{6}
\end{equation}
preicsely matches the weights corresponding to the non-perturbative effects we obtain from the recursion relations.

We pause to emphasize that this analysis does \emph{not} imply that operators with negative conformal weights actually appear in the OPE of local operators in a unitary holographic CFT, nor that conical excess geometries appear in the sum over geometries in the semiclassical gravitational path integral of AdS$_3$ gravity. Rather, the odd conical excesses should be thought of as solutions of the equations of motion that do not lie on the contour of integration in the gravitational path integral, but that nevertheless control the asymptotics of the $1/c$ expansion. This can be viewed as a sharp constraint on the sum over geometries in the gravitational path integral --- the sum over non-perturbative effects must conspire to reproduce the asymptotic behaviour discovered in section \ref{sec3}.

Of course, the actual mechanism by which this occurs will depend on the details of the bulk theory of gravity, or correspondingly of the large-$c$ conformal field theory.  In general, there may be many UV theories which in their low energy limit include three dimensional Einstein gravity.  Similarly, there are presumably many conformal field theories with large central charge.  Knowledge of the vacuum block alone is not enough to completely determine the theory, absent some other (presumably very stringent) constraints.  Reflecting this, we expect there will be many ways that the Borel transform of the vacuum conformal block can be completed to form a consistent CFT, and a corresponding theory of gravity in AdS.
What the have shown, however, is that any mechanism which accomplishes this {\it must} resolve the singularities in the Borel plane which correspond to the operators (\ref{eq:excessWeight}).   It still remains an important open question to determine exactly how these constraints can be most efficiently implemented in order to carve out the space of large$-c$ CFTs.  

\subsection{Chern-Simons quantization}
\label{sec:csquant}
Three-dimensional gravity with negative cosmological constant is famously related to $PSL(2,\mathbb{R})\times PSL(2,\mb{R})$ Chern-Simons theory \cite{Witten:1988hc}.  It turns out that this leads to an elegant reformulation of the result described above. 
We will begin by reviewing the (perturbative) relationship between 3d gravity and Chern-Simons theory, which will contextualize our finding that conical excess geometries control the asymptotic $1/c$ expansion of Virasoro conformal blocks and non-perturbative effects in three-dimensional gravity.\footnote{We are very grateful to Lorenz Eberhardt for discussions related to the material in this subsection.}

\subsubsection*{Relationship between Chern-Simons and 3D gravity} In the first-order formalism, three-dimensional gravity admits a description in terms of a vielbein $e_{\mu}^{a}$ and a spin connection $\omega_{\mu}^{ab}$.\footnote{Here, Greek indices $\mu,\nu,\ldots$ are meant to represent spacetime indices while Roman indices $a,b,\ldots$ denote local Lorentz indices.} The Einstein-Hilbert action on a three-manifold $M$ with negative cosmological constant $\Lambda = -{1\over \ell^2}$ can be written as\footnote{We are neglecting potential boundary terms here.} 
\begin{equation}\label{eq:IEH}
  I_{\rm EH}[e,\omega] = {1\over 16\pi G_N}\int_M d^3 x\,\epsilon^{\mu\nu\rho}\epsilon_{abc}\left[e^c_\rho\left(\partial_\mu \omega_\nu^{ab}-\partial_\nu\omega_\mu^{ab} + [\omega_\mu,\omega_\nu]^{ab}\right) + {1\over 3\ell^2}e_\mu^a e_\nu^b e_\rho^c\right]
\end{equation}
It is convenient to repackage the dualized spin connection $\omega_\mu^a = \half\epsilon^{abc}\omega_{\mu b c}$ and vielbein into $\mathfrak{psl}(2,\mb{R})$-valued one-forms $\mc{A}^+$ and $\mc{A}^-$
\begin{equation}\label{eq:GaugeFieldVielbein}
  \mc{A}^{\pm a}\equiv\left(\omega_\mu^a \pm {1\over \ell}e_\mu^a\right)dx^\mu.
\end{equation}
In this language, the Einstein-Hilbert action (\ref{eq:IEH}) can be rewritten as the action of $PSL(2,\mb{R})\times PSL(2,\mb{R})$ Chern-Simons theory
\begin{equation}\label{eq:IEHCS}
  I_{\rm EH}[e,\omega] = k I_{\rm CS}[\mc{A}^+] - kI_{\rm CS}[\mc{A}^-],
\end{equation} 
where as usual 
\begin{equation}
  I_{\rm CS}[\mc{A}] = {1\over 4\pi}\int_M \, \tr\left(\mc{A}d\mc{A} + {2\over 3}\mc{A}\wedge \mc{A}\wedge d \mc{A}\right).
\end{equation}
The level $k$ is related to the AdS length $\ell$ via
\begin{equation}
  k = {\ell\over 4 G_N} = {c\over 6},
\end{equation}
where $c = {3\ell\over 2 G_N}$ is the Brown-Henneaux central charge \cite{Brown:1986nw}. In order to discuss the phase space 
of the theory, we will consider the theory on $M=\Sigma\times \mb{R}$, where $\Sigma$, a compact Riemann surface, is the initial value surface. Then the constraint equations derived from (\ref{eq:IEH}) tell us that the classical phase space of $(2+1)$-dimensional gravity is given by the moduli space of flat $PSL(2,\mb{R})\times PSL(2,\mb{R})$ connections 
on $\Sigma$ modulo gauge transformations.

We pause to clarify a potentially confusing point. Although the previous discussion of the phase space quantization of Chern-Simons theory is inherently Lorentzian (indeed, we are considering $PSL(2,\mb{R})\times PSL(2,\mb{R})$ rather than $PSL(2,\mb{C})$ Chern-Simons theory), we will ultimately use these considerations to define the Hilbert space of the gravitational theory on (Euclidean) hyperbolic three-manifolds with asymptotic boundaries. Mathematically, the reason that this can be done is that the phase space should -- strictly speaking -- be the cotangent bundle of the moduli space of flat $PSL(2,{\mathbb R})$ connections, but this bundle is, due to a theorem of Bers, equivalent to two copies of of this moduli space (see e.g. \cite{Scarinci:2011np,Kim:2015qoa, Maloney:2015ina}).  Physically, this a version of the Hartle-Hawking construction of the Hilbert space of three dimensional gravity.  Namely, the Hilbert space is constructed by specifying boundary conditions at the asymptotic boundaries in Euclidean signature, rather than at fixed time. 

\subsubsection*{The moduli space of flat connections and Teichm\"uller space} The relationship between the classical phase space of AdS$_3$ gravity and the moduli space of flat $PSL(2,\mb{R})$ connections is more subtle than the above discussion might suggest (see \cite{Eberhardt:2022wlc} for a nice recent discussion). The moduli space of flat $PSL(2,\mb{R})$ connections on a Riemann surface $\Sigma$ in general has several disconnected components.  The classical phase space of 3D gravity on the other hand is connected and, as we will see, corresponds to a particular component of this moduli space. 

The space of flat $PSL(2,\mb{R})$ bundles on $\Sigma$ 
is parameterized by monodromies of the gauge field, in other words, by homomorphisms $\phi$ that map elements of the fundamental group of the Riemann surface $\pi_1(\Sigma)$ to $PSL(2,\mb{R})$ (modulo an overall conjugation by $PSL(2,\mb{R})$). If $\Sigma$ is a Riemann surface of genus $g>0$, then this space has $4g-3$ disconnected components.
The different components are labelled by an integer topological invariant known as the \emph{Euler class} $e$. The absolute value of the Euler class is known to be bounded by the Euler characteristic of $\Sigma$ \cite{Milnor_1958,Wood_1971,Goldman_1988}
\begin{equation}\label{eq:MilnorWood}
  |e| \leq 2g-2.
\end{equation}
The equality $|e| = 2g-2$ is saturated if and only if $\phi$ is an isomorphism of the fundamental group onto a discrete subgroup $\Gamma\subset PSL(2,\mb{R})$ \cite{Goldman_1988}.\footnote{Such $\Gamma$ are sometimes referred to as ``Fuchsian'' subgroups of $PSL(2,\mb{R}$).}

Let us briefly elaborate on the latter statement. A Riemann surface $\Sigma$ is parameterized by a moduli space of complex structures. The Teichm\"uller space $\mc{T}_\Sigma$ parameterizes the space of complex structures on $\Sigma$ up to infinitesimal diffeomorphisms. A Riemann surface $\Sigma$ with $g>1$ admits a unique hyperbolic structure. We may realize every such surface as a quotient of the upper half-plane by a Fuchsian subgroup $\Gamma\subset PSL(2,\mb{R})$
\begin{equation}
  \Sigma = \mb{H}^2/\Gamma
\end{equation}
with $\Gamma\simeq \pi_1(\Sigma)$. The subgroup $\Gamma$ may be specified by a homomorphism $\phi$ that maps the fundamental group $\pi_1(\Sigma)$ to $\Gamma\subset PSL(2,\mb{R})$ up to overall conjugation by $PSL(2,\mb{R})$. These are precisely the homomorphisms described at the end of the previous paragraph. The upshot of this discussion is that the component of the phase space of flat $PSL(2,\mb{R})$ connections on $\Sigma$ with maximal Euler class $|e| = 2g-2$ is isomorphic to the Teichm\"uller space of the Riemann surface. Hence the gravitational phase space on $\Sigma\times \mb{R}$ should really be regarded as the  ``Teichm\"uller component'' of the moduli space of flat $PSL(2,\mb{R})$ connections, and is thus given by two copies of the Teichm\"uller space $\mc{T}_\Sigma$ prior to gauging large diffeomorphisms on $\Sigma$ \cite{Krasnov:2005dm,Scarinci:2011np}.

\subsubsection*{The other components of the phase space of flat connections and conical excesses}
In order to describe the other disconnected components of the phase space of flat $PSL(2,\mb{R})$ bundles we must allow hyperbolic structures on $\Sigma$ with singularities. In particular, we allow coordinate charts that at isolated points are of the form $z\mapsto z^k$, $k\in\mb{Z}_{>1}$. 
Such a chart defines a conical singularity of angle $\theta = 2\pi k$. In a case with $m$ such singular points the Euler class is given by \cite{MR957518}
\begin{equation}
  e = 2-2g + \sum_{i=1}^m\left({\theta_i\over 2\pi}-1\right).
\end{equation}
In order to establish the inequality (\ref{eq:MilnorWood}), one appeals to uniformization theorems concerning singular hyperbolic structures, 
which show that for a given Riemann surface $\Sigma$ with $m$ isolated singular points as above, there exists a unique singular hyperbolic structure provided
\begin{equation}
  2-2g + \sum_{i=1}^m\left({\theta_i\over 2\pi} - 1\right) \leq 0.
\end{equation}
The cases with positive Euler class are related to those with negative Euler class by orientation reversal. Since this discussion is a bit formal and the role of the Euler class in 3D gravity may be unfamiliar, in appendix \ref{app:eulerclassstuff} we give a simple explicit computation which demonstrates that the conical excess singularities under consideration change the Euler class by an integer. 

So we see that the other, non-Teichm\"uller components of the moduli space of flat $PSL(2,\mb{R})$ connections on $\Sigma$ with sub-maximal Euler class are associated with hyperbolic structures on $\Sigma$ in the presence of conical excess singularities.  Of course, it is not an accident that it is precisely these conical excesses that appear as Borel singularities for vacuum blocks, as we will see.

\subsubsection*{Quantization of Teichm\"uller space} 

The appearance of conical excesses in both $PSL(2,{\mathbb R})$ Chern-Simons theory and in the Borel resummation of vacuum conformal blocks is not a coincidence.  The point is that the vacuum conformal block is what we get from quantizing three dimensional gravity perturbatively around a given background, and in particular describes the contribution from the quantization of the Teichm\"uller component of moduli space.   

In order to quantize the gravitational phase space we need to discuss the quantization of the Teichm\"uller space $\mc{T}_\Sigma$ associated with the spatial slices. The quantization of Teichm\"uller space was first carried out by H. Verlinde in \cite{Verlinde:1989ua} and later refined by Kashaev and Teschner \cite{Kashaev:1998fc,Teschner:2003at,Teschner:2003em,Teschner:2005bz}.
In Verlinde's paper, it was shown that with a non-standard choice of polarization, the flatness conditions, translated to constraints on the quantum wavefunction, are equivalent to the Virasoro Ward identities. The conclusion is that the quantum Hilbert space of three-dimensional gravity is given by \cite{Verlinde:1989ua,Kim:2015qoa}
\begin{equation}
  \mc{H}_{\rm grav} = \mc{H}_{\Sigma}\otimes \overline{\mc{H}}_{\Sigma},
\end{equation}
where $\mathcal{H}_\Sigma$ is the space of holomorphic Virasoro conformal blocks on $\Sigma$ with all intermediate conformal weights above threshold, $h_i \geq {c-1\over 24}$. 
The constraint that the intermediate weights of the conformal blocks lie above the $c-1\over 24$ threshold is a consequence of normalizability of the wavefunctions in the Hilbert space of AdS$_3$ gravity. 
The normalizability constraint was emphasized in \cite{Kim:2015qoa}, which considered the quantization of the Teichm\"uller space in the case important for holography that $\Sigma$ has a single connected boundary.
Indeed, the Hilbert space is equipped with an explicit inner product between conformal blocks \cite{Verlinde:1989ua,Kim:2015qoa} (the precise form of this inner product is not relevant for our present purposes). Conformal blocks with internal weights below the $c-1\over 24$ threshold, particularly the identity block that we study throughout this paper, correspond to non-normalizable wavefunctions in quantum Teichm\"uller theory \cite{Kim:2015qoa}. The (delta-function) normalizable wavefunctions are precisely the conformal blocks that appear in the decomposition of correlation functions in the non-compact Liouville CFT, and indeed the Liouville CFT and quantum Teichm\"uller theory are intertwined in a way akin to the relationship between WZW models and Chern-Simons theories based on a compact gauge group.\footnote{The relationship between the TQFT defined by the quantization of the ``Teichm\"uller component'' and 3d gravity will be explored in much more detail in \cite{TTQFTPaper}.} The quantization of the non-Teichm\"uller components, on the other hand, is not well understood.

\subsubsection*{The mapping class group} 

In gravity one must gauge not only the infinitesimal diffeomorphisms but also the large diffeomorphisms not continuously connected to the identity. For a Riemann surface $\Sigma$, the large diffeomorphisms are parameterized by the mapping class group\footnote{Here $\Diff_0(\Sigma)$ is meant to denote the diffeomorphisms continuously connected to the identity.}
\begin{equation}
  \MCG(\Sigma) = \Diff(\Sigma)/\Diff_0(\Sigma).
\end{equation}
As a result the phase space $\mc{M}_{{\rm grav},\Sigma}$ of $(2+1)$-dimensional gravity on $\Sigma\times \mb{R}$ inherited from the identification (\ref{eq:GaugeFieldVielbein}) is identified with the product of Teichm\"uller spaces modded out by the action of the diagonal mapping class group 
\begin{equation}
  \mc{M}_{{\rm grav},\Sigma} \simeq (\mc{T}_\Sigma\times\mc{T}_\Sigma)/\MCG(\Sigma).
\end{equation}
So at least in perturbation theory, it is the gauging by the mapping class group that distinguishes $(2+1)$-dimensional gravity from gauge theory.

To summarize, we have seen that three-dimensional gravity with negative cosmological constant admits a classical reformulation in terms of $PSL(2,\mb{R})\times PSL(2,\mb{R})$ Chern-Simons theory. The classical phase space of this Chern-Simons theory on $\Sigma\times \mb{R}$ is the moduli space of flat $PSL(2,\mb{R})$ connections on $\Sigma$ (modulo gauge transformations), which in turn has multiple disconnected components classified by the Euler class $e$. The component with maximal Euler class $e = \pm|\chi(\Sigma)|$ is equivalent to the Teichm\"uller space of hyperbolic structures on $\Sigma$, and within this component the associated vielbein is invertible and the metric is 
completely well-defined. Before gauging large diffeomorphisms, the gravitational phase space is then given by two copies of Teichm\"uller space. The other components are associated with hyperbolic structures on $\Sigma$ in the presence of isolated conical excess singularities. 

Our conclusion is that the conical excess geometries which control the asymptotics of the $1/c$ expansion of Virasoro conformal blocks 
are a signature of tunnelling between the Teichm\"uller component and the other components of the phase space of flat $PSL(2,\mb{R})$ connections on $\Sigma$.  This implies that we cannot obtain a consistent quantum theory with a smooth classical limit just by quantizing the space of non-singular metrics, i.e. the Teichm\"uller component.  In other words, {\it quantum general relativity cannot be a theory of perturbative gravitons alone}.  It must include something else.  In the simple case of $PSL(2,{\mathbb R})$ Chern-Simons theory, we can add the other components of moduli space to cure this problem.  But these would violate unitarity in a theory of gravity with positive Newton's constant.  So we must add heavy states of positive dimension.  In the gravitational language, these are most naturally interpreted as black hole microstates.

\subsection{Conical excesses and the monodromy method for Virasoro conformal blocks}

In this section we interpret these results using the monodromy method to describe conformal blocks in the large $c$ limit.
In particular, we have already seen that the asymptotic structure of the vacuum block can be interpreted in terms of novel subleading ``saddle point" contributions to the original four point function.  It turns out that these other saddle points can be viewed as formal solutions of the same monodromy problem used to compute the original vacuum block in the large $c$ limit. Before proceeding we note that a very similar observation was already made in section 2.1 of \cite{Fitzpatrick:2016mjq}.

To see this, let us briefly recall the monodromy technique used to computed Virasoro conformal blocks in the $c\to\infty$ limit, when both the external and internal operator weights are of order $c$  (we refer to \cite{Hartman:2013mia,Faulkner:2013yia,Harlow:2011ny,Fitzpatrick:2014vua} for more thorough recent reviews). The idea is to consider the five-point function
\begin{equation}
    \langle O_{1}(x_{1})O_{2}(x_{2}) \phi_{2,1}(z) O_{3}(x_{3}) O_{4}(x_{4}) \rangle
\end{equation}
with the insertion of the degenerate field $\phi_{2,1}$ with dimension $h_{2,1} = -(3b^2 + 2)/4$ where $c = 13 + 6\left( b^2 + b^{-2} \right)$. 
In the limit $b\rightarrow{0}$, $h_{2,1}$ does not scale with $c$. So at large $c$ one can use the fact that the semiclassical block exponentiates \cite{Besken:2019jyw} to argue 
that this five point function is proportional to our original four point function times a wave-function  $\psi(z,x_{i})$ associated with insertion of the field $\phi_{2,1}$: 
\begin{equation}\label{meq}
    \begin{split}
    \langle O_{1}(x_{1})O_{2}(x_{2}) \phi_{2,1}(z) O_{3}(x_{3}) O_{4}(x_{4}) \rangle 
   &= \sum_{p, \{k\} } \langle O_{1}O_{2} \ket{O_{p}^{ \{k\} }} \bra{O_{p}^{ \{k\} }} \phi_{2,1}(z)\, O_{3} O_{4} \rangle \\
   &= 
   \psi(z,x_{i}) \sum_{p} C_{12p}C_{34p}\, F_{h_{p}}(x_{i}) \bar{F}_{\bar{h}_{p}}(x_{i}) \\
   \end{split}
\end{equation}
The Ward identities for the degenerate field imply that the five point function satisfies a second-order differential equation
\begin{equation}
    \label{diffeq}
    \left( -\frac{3}{2(2h_{2,1}+1)}\partial^{2}_{z} + \sum_{i=1}^{4}\left( \frac{h_{i}}{(z-x)^{2}} + \frac{1}{z-x_{i}}\partial_{i} \right) \right)\langle O_{1}(x_{1})O_{2}(x_{2}) \phi_{2,1}(z) O_{3}(x_{3}) O_{4}(x_{4}) \rangle = 0
\end{equation}
Each term in the sum (\ref{meq}) must satisfy this same differential equation, so at large $c$ we can substitute $\psi(z,x_{i})F_{h_{p}} (x_{i})= \psi(z,x_{i})e^{-\frac{c}{6}f(x_{i})}$ to obtain
\begin{equation}
    \label{mondeq}
    \partial_{z}^{2} \psi(z) + T(z) \psi(z) = 0
\end{equation}
\begin{equation}
    \begin{split}
    T(z) = \frac{\epsilon_{1}}{z^{2}} + \frac{\epsilon_2}{(z-x)^2} +& \frac{\epsilon_3}{(z-1)^2} + \frac{\epsilon_1 + \epsilon_2 + \epsilon_3 - \epsilon_4}{z(1-z)} + \frac{c_{2}(x) x(1-x)}{z(z-x)(z-1)} \\ 
    &\epsilon_{i} = \frac{6h_{i}}{c}  \,, \quad c_{2}(x) = \frac{\partial f}{\partial x}
    \end{split}
\end{equation}
where using the conformal transformation we have set the insertion points to be $x_1 = 0, x_2 = x, x_3 = 1, x_4 = \infty$ and in (\ref{mondeq}) we suppress the dependence on $x$.

Let us now consider the three-point function
\begin{equation}
    \label{shorteq}
    \langle O_{\alpha}(x_1) \phi_{2,1}(x_2) O_{\beta}(x_3) \rangle = \frac{C_{\alpha \phi \beta}}{ x_{12}^{ h_{\alpha} + h_{2,1} - h_{\beta} } x_{13}^{ h_{\alpha} + h_{\beta} - h_{2,1} }  x_{23}^{ h_{2,1} + h_{\beta} - h_{\alpha}}  }
\end{equation}
This must satisfy a differential equation similar to (\ref{diffeq})
\begin{equation}
    \left( -\frac{3}{2(2h_{2,1}+1)}\partial^{2}_{3} + \sum_{i=1,2} \left( \frac{h_{i}}{(z-x)^{2}} + \frac{1}{z-x_{i}}\partial_{i} \right) \right) \langle O_{\alpha}(x_1) \phi_{2,1}(x_2) O_{\beta}(x_3) \rangle = 0
\end{equation}
Substituting (\ref{shorteq}), one finds that the differential equation is satisfied when 
\begin{equation}
    \label{opecon}
    h_{\beta} - h_{\alpha} - h_{2,1} = \frac{1}{2}\left( 1 \pm \sqrt{1 - \frac{24 h_{\alpha}}{c} } \right) \,, \quad c\gg1
\end{equation}
Going back now to the five-point function we saw that in the conformal block decomposition we have 
\begin{equation}
    \langle O_{1}O_{2}\, \phi_{2,1}(z)\, O_{3} O_{4} \rangle = \sum_{\alpha}\sum_{\beta} C_{12\alpha} C_{34\beta}\, \langle O_{\alpha} \phi_{2,1}(z) O_{\beta} \rangle
\end{equation}
where we have done an OPE expansion of $O_{1} O_{2}$ and $O_{3} O_{4}$. But from the discussion above we know that $h_{\beta}$ must satisfy (\ref{opecon}). So as $z$ encircles the insertion point of $O_{\alpha}$ (alternatively around $x_{1}$ and $x_{2}$) we get the monodromy matrix
\begin{equation}\label{mmatrix}
    M_h = \begin{pmatrix} e^{\pi i\left(1+\sqrt{1-{24 h_{\alpha}\over c}}\right)} & 0 \\ 0 & e^{\pi i\left(1-\sqrt{1-{24h_{\alpha} \over c}}\right)}\end{pmatrix}
\end{equation}
as desired. The monodromy matrix is fixed by the dimension of the internal primary.

The monodromy matrix evaluates to the identity when the intermediate primary is the identity operator $(h=0)$. We get the same matrix for other non-zero $h$ when it satisfies
\begin{equation}
    \begin{split}
    &\sqrt{1-{24 h\over c}} = r \,,\quad r\in\mathbb{Z}_{odd} \\
    &h = \frac{c}{24}(1-r^{2}) \\
    \end{split}
\end{equation}
which is precisely the dimension of the $\phi_{r,1}$ fields at large central charge.
Our conclusion is that the new saddle points we have discovered using our asymptotic series can be viewed as formal solutions of the same monodromy problem used to compute the original vacuum block in the large $c$ limit.

How, then, is this related to the analysis of conformal blocks which arises from the quantization of Teichm\"uller space?  
The point is that the solutions of the monodromy problem are the semi-classical WKB contributions to the conformal block, viewed as a wave function on Teichm\"uller space.  The usual solution of the monodromy problem gives the leading conformal block.  The other solutions simply provide the ``tunnelling" solutions described in the previous section.

\section{Discussion and future directions}
\label{sec:discuss}

One of our main results is a new way of answering the question: is there a quantum theory of gravity which includes only gravitons and nothing else?  At least in three dimensions, the answer is a resounding no: such a theory is not well defined on its own, as it requires non-perturbative effects in order to render graviton perturbation theory finite.  This is perhaps an expected result, but it is gratifying that we can demonstrate it explicitly.
The next step, of course, is to use these considerations to constrain and understand these non-perturbative effects.  At negative $c$, there is a clear and unambiguous story.  But for unitary theories at positive $c$ much still remains to be understood.
There must be delicate contributions of heavy states that combine in order to cure the perturbative divergence of the vacuum block. In the dual gravitational picture, these heavy states are of course black holes.  
In this way we see that the perturbative structure of the graviton loop expansion is connected -- via resurgence -- to black hole physics. 
For example, although the conical excess ``saddles'' that we have identified from an asymptotic analysis of the Virasoro vacuum block must not lie on the physical contour of integration in the Euclidean gravitational path integral, we expect that they admit an interesting interpretation upon analytic continuation to Lorentzian signature in terms of late-time black hole physics; see \cite{Fitzpatrick:2016ive,Fitzpatrick:2016mjq,Chen:2016cms,Chen:2017yze} for some comments and results in this direction. In particular, the conical excess saddles were argued for (from different considerations) in \cite{Fitzpatrick:2016mjq} and in \cite{Chen:2016cms} were shown to be important for the late Lorentzian time behaviour of the out-of-time-ordered correlator expected of a maximally-chaotic theory.

We note that we have focused so far on resurgence for vacuum conformal blocks for four point functions, but the analysis for other blocks will proceed similarly.  Vacuum blocks for other observables, such as higher point functions and higher genus partition functions, are amenable to analysis using Zamolodchikov recursion relations.  So a generalization of the analysis presented in section \ref{sec3} implies that these blocks will have similar poles in the Borel plane corresponding to the exchange of odd integer conical excesses.  

Non-vacuum blocks should likewise have an asymptotic series in $1/c$, although the location of the poles in the Borel plane will be different.  In particular, if we consider the exchange of the conformal block for a primary operator of dimension $h$, we conjecture that there will be poles in the Borel plane corresponding to the exchange of operators of dimension $h_{n}$, where
\be
\sqrt{1-{24 h_{n}\over c}} = \sqrt{1-{24 h\over c}} + 2n 
\ee
for integer $n$.
This condition follows from requiring that $h_n$ has the same monodromy as the original operator $h$, according to (\ref{mmatrix}).
This can be rewritten suggestively by letting $h=\frac{c}{24}\left(1+p^2\right)$, where $p$ is a (rescaled) Liouville momentum at large $c$.  This means that the new poles in the Borel plane are related to the original operator by shifts $p_n = p + 2 i n$ of the rescaled Liouville momentum.  
Indeed, Liouville theory is a natural setting where these resurgence phenomena can be studied explicitly, and it is possible to give physical interpretations to these new saddle points, both in the language of Liouville theory and in 3d gravity.  This will be discussed in \cite{brilliantpaper}.

\section*{Acknowledgements}

We are very grateful to S. Caron-Huot, K. Dasgupta, L. Eberhardt, A. L. Fitzpatrick, M. Mari\~no, H. Maxfield, S. Shenker, E. Verlinde, H. Verlinde, and E. Witten for useful conversations. We thank A. L. Fitzpatrick for very helpful comments on a draft.
A.M. is supported in part by the Simons Foundation Grant No. 385602 and the Natural Sciences and Engineering Research Council of Canada (NSERC), funding reference number SAPIN/00047-2020.  N.B. is supported by the Sherman Fairchild Foundation and the U.S. Department of Energy, Office of Science, Office of High Energy Physics Award Number DE-SC0011632. The work of S.C. is supported by the Sam B. Treiman fellowship at the Princeton Center for Theoretical Science.
This work was performed in part at the Aspen Center for Physics, which is supported by National Science Foundation grant PHY-1607611.

\appendix

\section{Numerics}
\label{sec:numerics}
In this appendix, we follow the steps outlined in section \ref{sec3} for the numerical computation of the conformal blocks. For this we will use the code found in \cite{Chen:2017yze} which uses the $h$-recursion relation to compute the conformal blocks. We let
\begin{equation}
    \langle O_{1}(0)O_{2}(z)O_{3}(1)O_{4}(\infty) \rangle = \sum_{h,\bar{h}} P_{h,\bar{h}} \abs{F(c,h_{i},h,z)}^{2}
\end{equation}
\begin{equation}
    F(c,h_{i},h,z) = (16q)^{h-\frac{c-1}{24}}z^{\frac{c-1}{24}-h_{1}-h_{2}}(1-z)^{\frac{c-1}{24}-h_{2}-h_{3}}\theta_{3}(q)^{\frac{c-1}{2}-4\sum_{i}h_{i}}H_{c,h_{i},h}(q)
\end{equation}
where
\begin{equation}
    q = e^{-\frac{\pi K(1-z)}{K(z)}} \,, \quad z = \left( \frac{\theta_{2}(q)}{\theta_{3}(q)} \right)^{4}
\end{equation}
The mathematica code in \cite{Chen:2017yze} can be used to compute the coefficients in the series expansion
\begin{equation}
    H_{c,h_{i},h}(q) = 1 + \sum_{n\ge1}c_{n}(c,h_{i},h)\, q^{n}
\end{equation}
where we have pairwise identical external operators. For the present work, we evaluate $H(q)$ up to a fixed order in $q^{J}$ and then compute $\ln H(q)$ to same order in $q$ expansion. We expect this to have an asymptotic expansion in $1/c$ at large $c$ with the coefficient of the order $c$ term contributing to the classical block.
\begin{equation}
    \ln H(q) = \sum_{n=-1}^{\infty} \frac{a_{n}(q)}{c^{n}}
\end{equation}
The leading non-perturbative effect in the $\ln H(q)$ is the same as in $H(q)$ and so we are allowed to take the logarithm. To justify this we write
\begin{equation}
    \begin{split}
    H(q) = \exp[\ln H(q)] &= \exp[c\,S_{cl} + \sum_{n\ge0}\frac{a_{n}}{c^{n}} + e^{-cS_{*}}(\dots) + \dots] \\
    &= \exp[c\,S_{cl} + \sum_{n\ge0}\frac{a_{n}}{c^{n}}] \exp[e^{-cS_{*}}(\dots) + \dots] \\
    &= \exp[c\,S_{cl} + \sum_{n\ge0}\frac{a_{n}}{c^{n}}] \left( 1 + e^{-cS_{*}}(\dots) + e^{-c\,2S_{*}}(\dots) + \dots \right) \\
    \end{split}
\end{equation}
so we see that even $H(q)$ has the leading non-perturbative contribution as $e^{-cS_{*}}$. In particular, the higher order terms can be interpreted as the contribution of multi-instanton terms. This can also be seen by growth of the coefficients in $1/c$ expansion. We have
\begin{equation}
    H(q) = e^{c\,S_{cl} + a_{0}}\sum_{n=1}^{\infty} \frac{b_{n}(q)}{c^{n}} \,,\quad \ln H(q) = \sum_{n=-1}^{\infty} \frac{a_{n}(q)}{c^{n}}.
\end{equation}
Here, the $b_{n}$ and $a_{n}$ are related as 
\begin{equation}
    b_{n} = a_{n} + \frac{1}{2!}\sum_{k=0}^{n}a_{k}a_{n-k} + \sum_{p=3}^{n}\frac{1}{p!}\sum_{k_{i}\in (n,p)}a_{k_{1}}a_{k_{2}}\dots a_{k_{p}}  
\end{equation}
where $(n,p)$ denotes the set of partition of $n$ into $p$ integers. If we assume factorial growth for these coefficients
\begin{equation}
    b_{n}\sim n!S_{1}^{-n} \,, \quad a_{n}\sim n!S_{2}^{-n}
\end{equation}
then we have
\begin{equation}
    S_{1}^{-n} \sim S_{2}^{-n} \left( 1 + \frac{1}{n!}\sum_{p=2}^{n}\frac{1}{p!}\sum_{k_{i}\in (n,p)}k_{1}!k_{2}!\dots k_{p}! \right)
\end{equation}
At large $n$ the sub-leading terms are exponentially suppressed. For instance, for $p=2$, consider the partition $n = n/2 + n/2$. 
\begin{equation}
    \frac{\frac{n}{2}!\frac{n}{2}!}{n!2} \sim \frac{1}{2^{n+1}}
\end{equation}
so at large $n$, we have $S_{1}=S_{2}$ and hence the leading non-perturbative terms agree for $H(q)$ and $\ln H(q)$. In particular, this sub-leading term gives $S_{1} = 2S_{2} $ which is like the two instanton term. This appeared above in the Taylor expansion of the exponential.

We now compute $\ln H(q)$ using the recursion relations and compute the coefficients in the $1/c$ expansion. Then we compute its Borel transform 
\begin{equation}
    \label{332}
    BZ(S) = \sum_{n\ge0}d_{n}S^{n} \,,\quad \left(d_{n}= \frac{a_{n}}{n!} \right)
\end{equation}
To find the radius of convergence of this series we can look at the limit
\begin{equation}
    \lim_{n\rightarrow\infty} \frac{d_{n}}{d_{n-1}} = \frac{1}{S_{*}}
\end{equation}
So, a plot of $\frac{d_{n}}{d_{n-1}}$ vs $\frac{1}{n}$ can be used to determine the radius of convergence by looking at the intercept on the $y$-axis. In practice, it is better to plot the coefficients \cite{doi:10.1137/0150091}
\begin{equation}
    \label{334}
    c_{n}^{2} = \frac{d_{n+1}d_{n-1}-d_{n}^{2}}{d_{n}d_{n-2}-d_{n-1}^{2}}
\end{equation}
if the pole in the Borel plane is a generic complex number. The intercept on the $y$-axis of the plot $c_{n}$ vs $\frac{1}{n}$ gives the reciprocal of the radius of the convergence $1/S_{*}$. As a check of the method, we can now compute the value of $S_{*}$ for the large $c$ expansion of the vacuum block appearing in the 4-point function of the degenerate fields $\phi_{2,1}$.

\subsection{Vacuum block $\langle \phi_{2,1}\phi_{2,1}\phi_{2,1}\phi_{2,1} \rangle$}

\subsubsection*{Exact Block:}
We have seen earlier how in this case we can compute the exact Borel transform for any $z$ since we have an one-dimensional contour integral representation.

\begin{equation}
    \begin{split}
    \langle \phi_{2,1}(\infty) \phi_{2,1}(1) \phi_{2,1}(z)  \phi_{2,1}(0) \rangle &=  \sum_{i=1,2} \frac{C_{i}^{2}}{z^{2h_{2,1}}\bar{z}^{2\bar{h}_{2,1}}} \, F(h_{i},z)\bar{F}(\bar{h}_{i},\bar{z}) \\
    \end{split}
\end{equation}
with the vacuum block given as
\begin{equation}
    \begin{split}
        \label{336}
        F(h_{i}=0,z) &= (1-z)^{-b^{2}/2} \int_{0}^{1} dx \, [x(1-x)(1-zx)]^{b^{2}} \\
        &= (1-z)^{-\frac{b^{2}}{2}}\,e^{b^{2}\psi(x_{-})} \int_{0}^{1} dx\, e^{b^{2}[ \psi(x) - \psi(x_{-})] }\\
        &= (1-z)^{-b^{2}/2}e^{b^{2}\psi(x_{-})} \int_{0}^{\infty} dS\, e^{-b^{2}S}\, BZ(S) \\
    \end{split}
\end{equation}
where $BZ(S)$ is simply the Jacobian in the change of coordinates from $x$ to $S$. Expanding $BZ(S)$ around $S=0$ gives the coefficients $d_{n}$ in (\ref{332}). Using this we compute the $c_{n}$ as defined in (\ref{334}) and plot them against $1/n$.

\begin{figure}[h]
    \centering
    \includegraphics[scale=0.35]{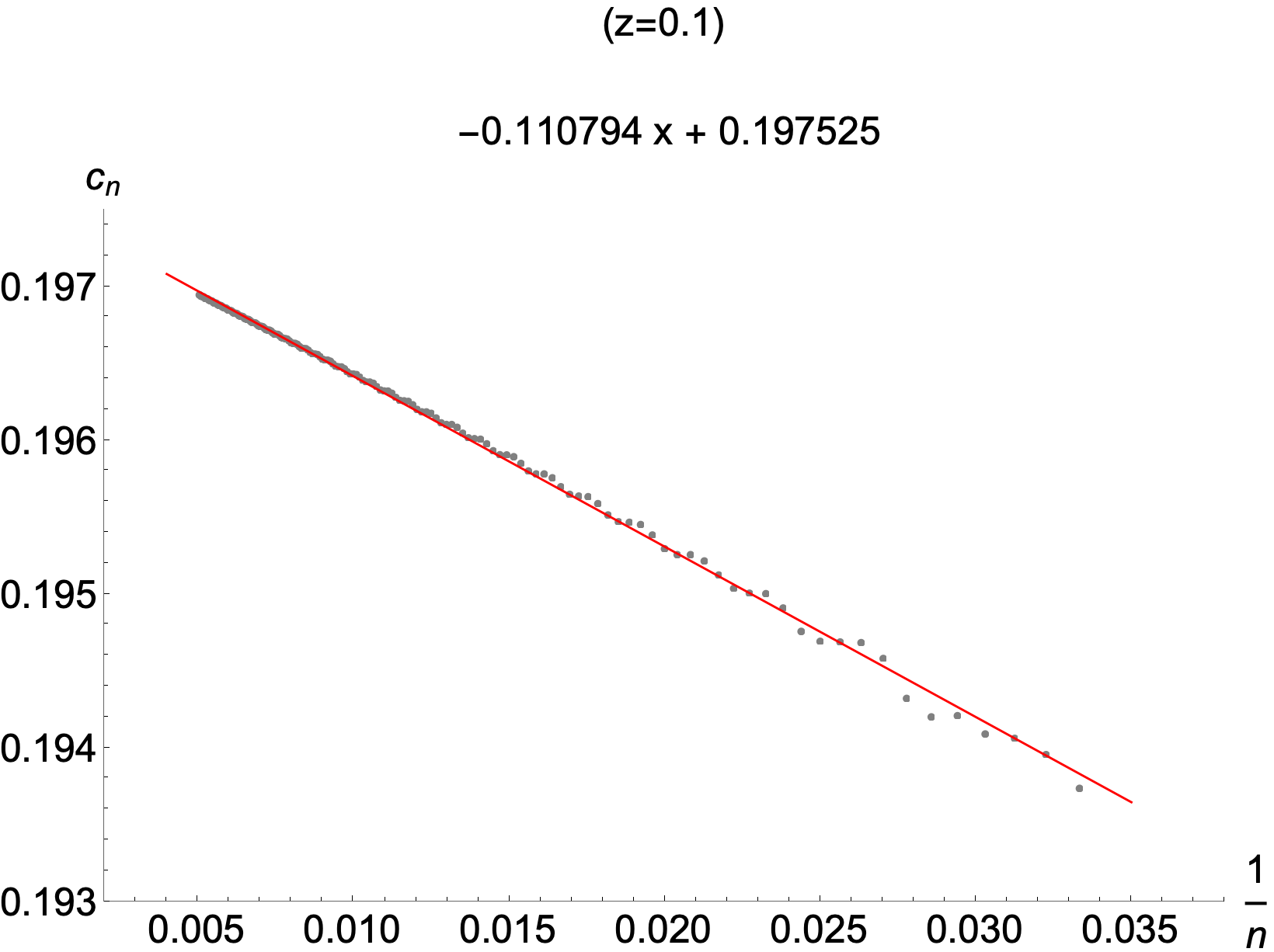} 
    \caption{The plot label denotes the value of cross-ratio $z$ and the equation of the linear fit to the curve. The value of the intercept gives the inverse of the radius of the convergence $1/S_{*}$.}
\end{figure}

The value of the intercept can also be predicted from the integral (\ref{336}).  It is the action of the second saddle point of that integral. Since we have defined $S(x) = \psi(x_{-}) - \psi(x)$, we have
\begin{equation}
     S(x_{+}) = \psi(x_{-}) - \psi(x_{+})
\end{equation}
For the cross ratio $z=0.1$, this is a complex number and to get the radius of convergence $S_*$, we take the absolute value.
\begin{equation}
    S_{*}^{-1}(z=0.1) = \abs{S(x_{+})}^{-1} = 0.197451 
\end{equation}
which agrees well with the value of the intercept that we find in the plot obtained by setting $x=0$ in the equation of the linear fit. 

\subsubsection*{Truncating in $z$}
We can also compute the vacuum block up to a fixed order in the $z$ expansion using the fact that the above integral for the vacuum block is a hypergeometric function
\begin{equation}
    F(h_{i}=0,z) \propto {}_{2}F_{1}(-b^{2},b^{2}+1,2b^{2}+2,z
    ) 
\end{equation}
This is good way to understand the effects of keeping only a finite number of terms in the $z$ expansion and then taking the large $c$ limit. This is important since when we use the recursion relations we will again be limited to finite order in the cross-ratio.

For the present purpose, we expand the logarithm of the hypergeometric function up to $\mathcal{O}(z^{30})$ and then write it as series in $1/b^{2}$. 
\begin{figure}[h]
    \centering
    \includegraphics[scale=0.27]{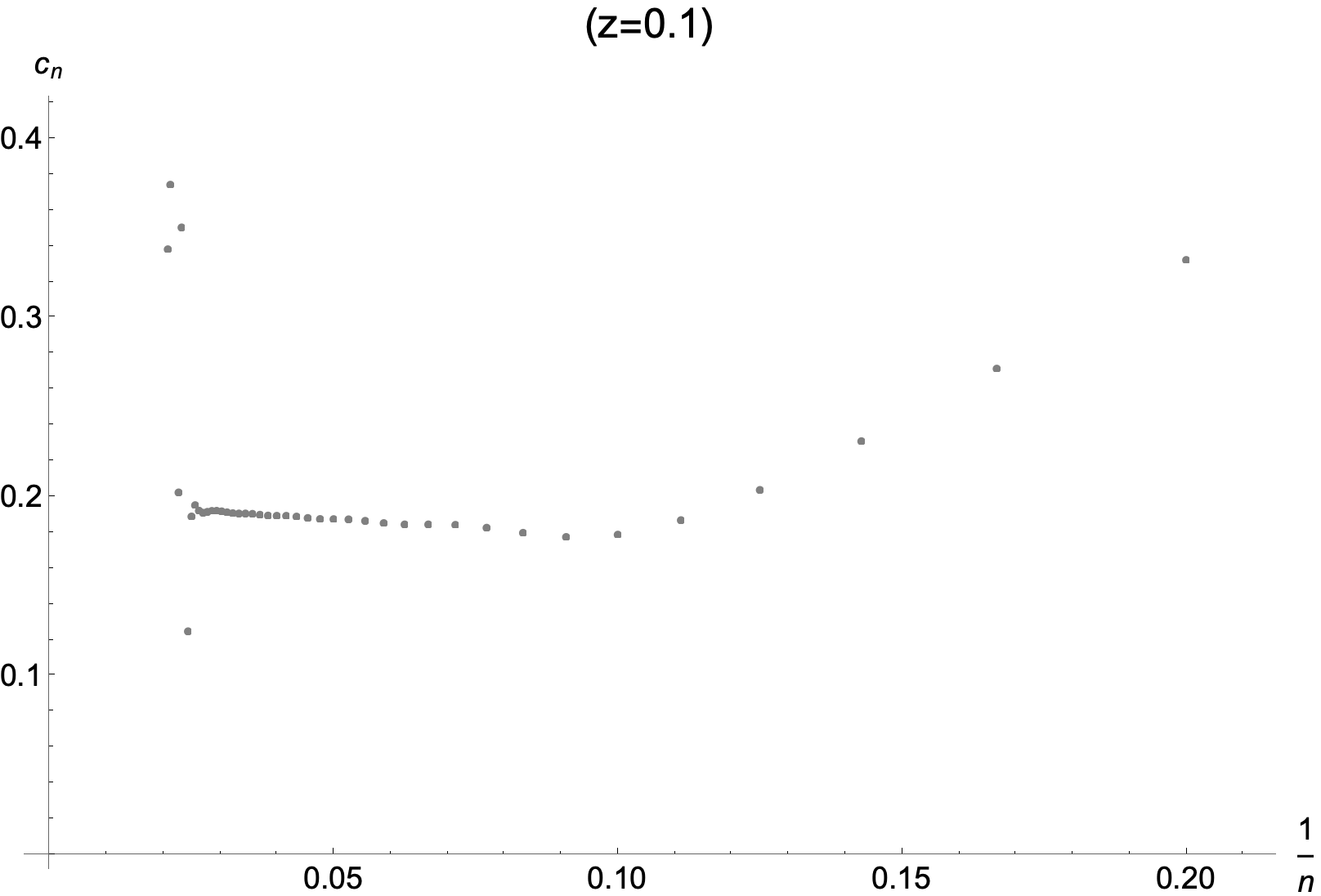}
    \hspace{1cm}
    \includegraphics[scale=0.27]{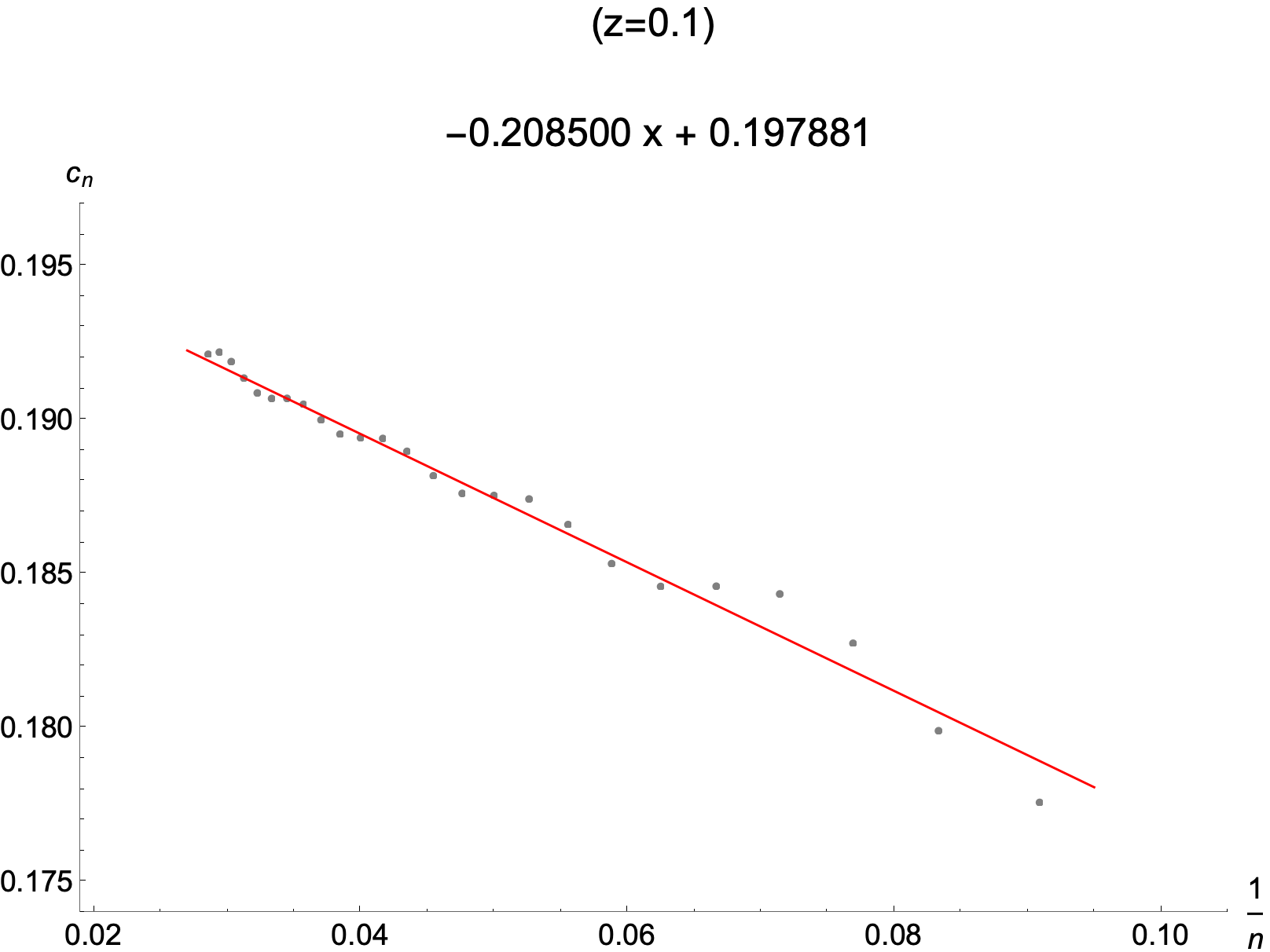}   
    \caption{$c_{n}$ vs $1/n$ for the truncated Hypergeometric function. Again, the value of the intercept gives the inverse of the radius of the convergence $1/S_{*}$.}
\end{figure}

In the first figure we see that, initially after a few data points the curve starts approaching a straight line but for higher values of $n$ it starts deviating. This is because of the fact that we have truncated the series at finite order in $z$ and so the higher order terms in the $1/b^{2}$ expansion are not reliable. In particular, numerically they have large coefficients and so we need more terms in the $z$ expansion.

However, in the first plot we can still zoom into the region where we have the straight line and then fit a linear curve. This is shown in the second plot. Again, the intercept agrees with the desired value.

\subsubsection*{Recursion relations}
Finally, we now use the recursion relations to repeat the above steps. This is important since for generic blocks we will not have any simple representation of the blocks and can only work to finite order in the cross-ratio expansion. 
We will be using the $h$-recursion relations mentioned earlier, and the code of \cite{Chen:2017yze}. Below, we plot similar quantities as in the previous two methods. 

\begin{figure}[h]
    \centering
    \includegraphics[scale=0.3]{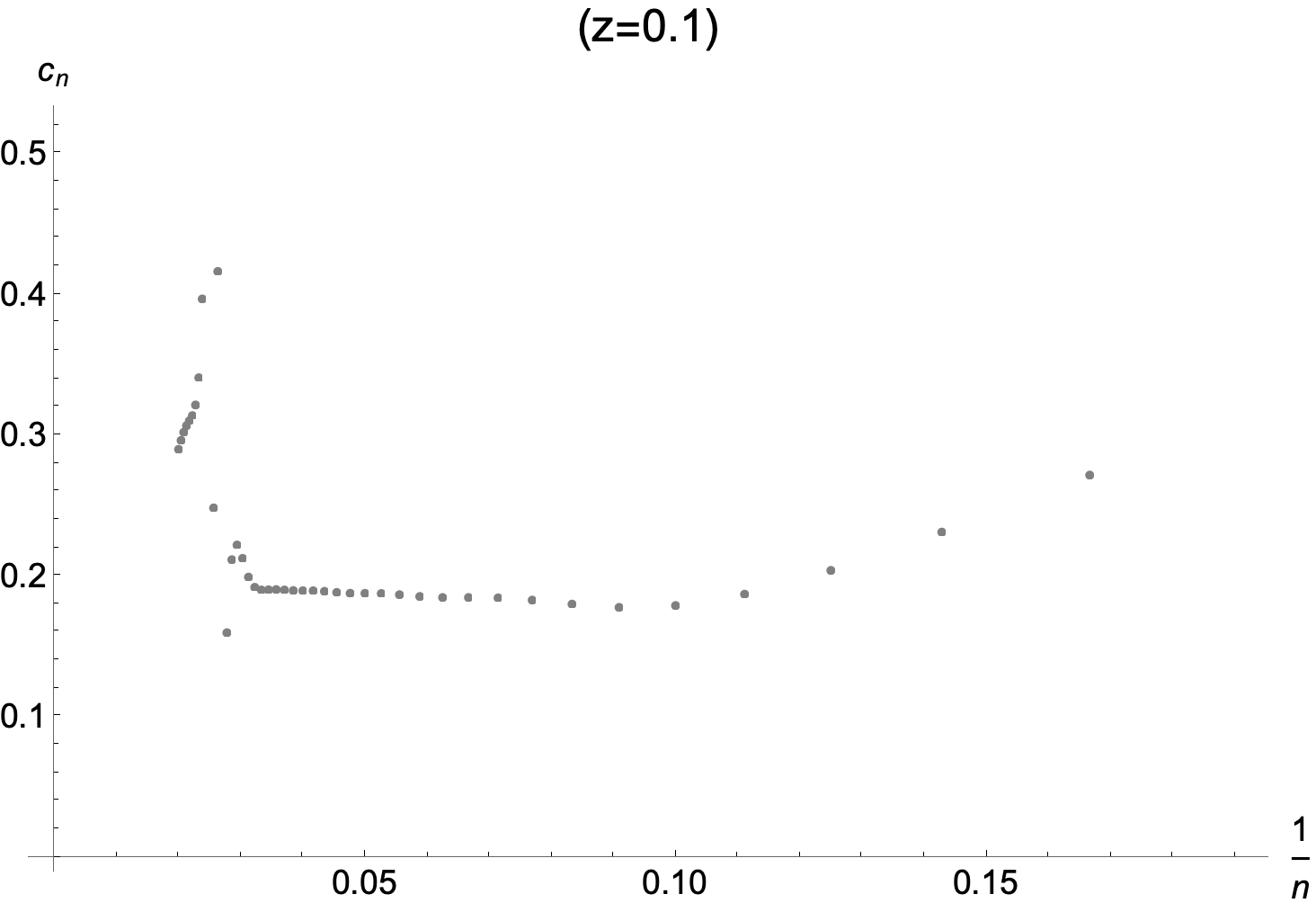} 
    \hspace{1cm}
    \includegraphics[scale=0.3]{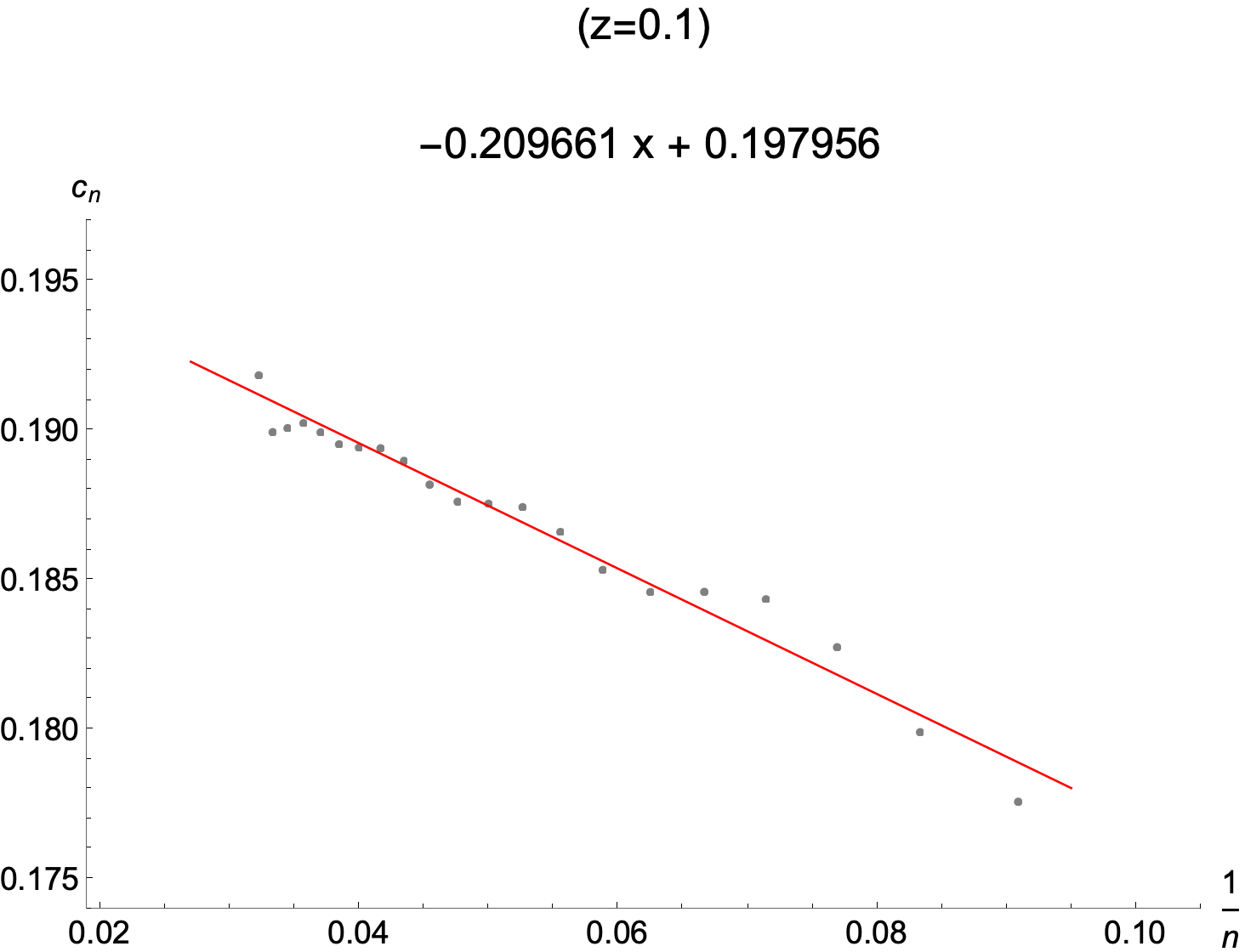} \\
    \caption{$c_{n}$ vs $1/n$. The recursion relation are solved to $\mathcal{O}(q^{26})$.}
\end{figure}
The linear fit gives an intercept close to the exact result of $S_{*}^{-1}(z=0.1) = 0.197451$. 

\subsection{All-light block $\langle \phi_{h_{L}}\phi_{h_{L}}\phi_{h_{L}}\phi_{h_{L}} \rangle$}
Here, we compute the vacuum conformal block appearing in the four point functions of primary operators with dimensions that do not scale with the central charge $c$ and are positive unlike the degenerate fields of minimal models i.e. $h_{L}/c \rightarrow 0$ as $c\rightarrow\infty$ and $h_{L}>0$. We will compute these blocks using the recursion relations mentioned above. We evaluate the coefficients to order $\mathcal{O}(q^{20})$ in the recursion relations.

\begin{itemize}
    \item $h_{L} = 10$
    \begin{figure}[h!]
        \centering
        \includegraphics[width=9.5cm]{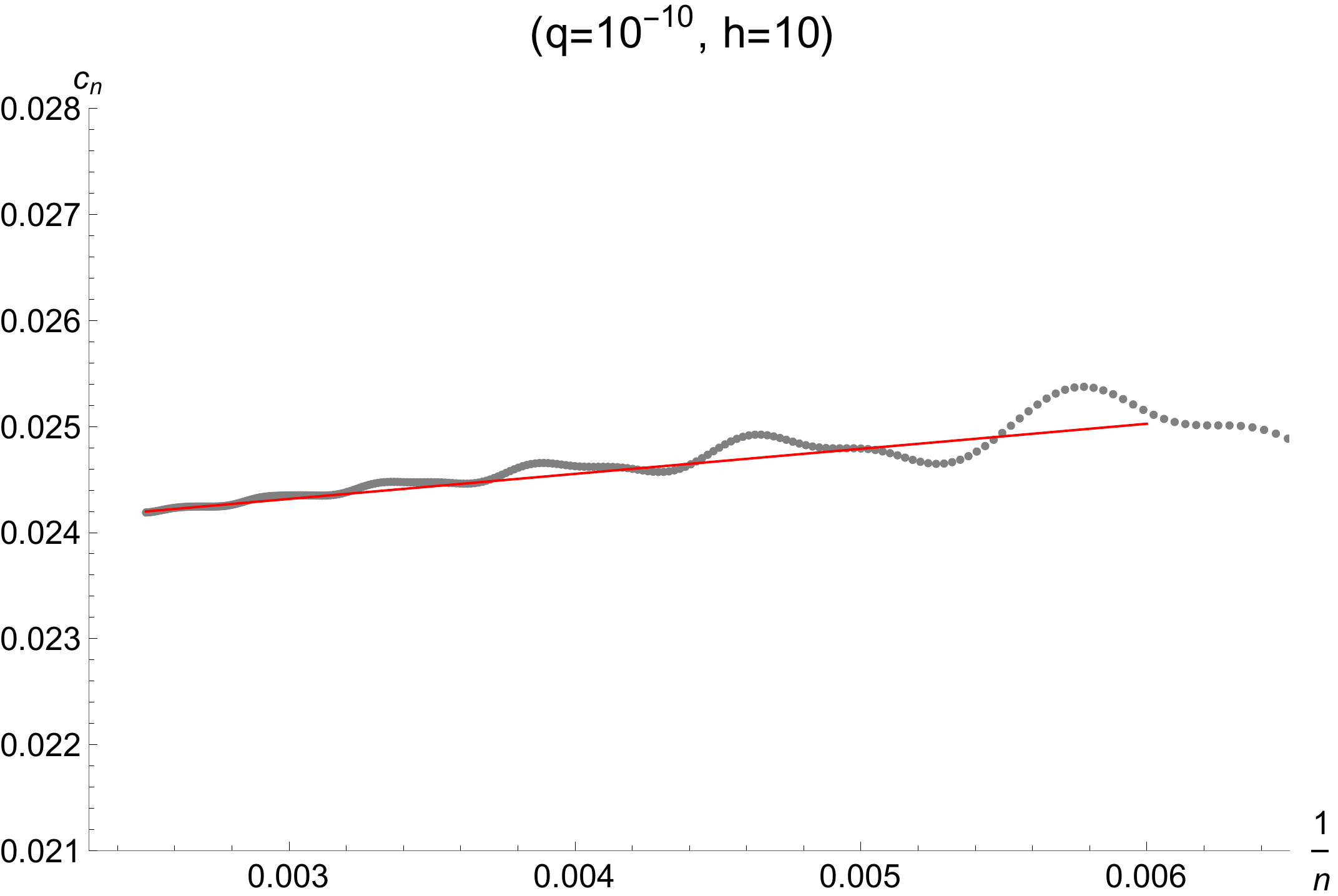} 
        \hspace{1cm} 
        \caption{$c_{n}$ vs $1/n$ for $h_{L}=10$.}
        \label{fig:hl10}
    \end{figure}

    The value of the intercept and the dimension of the primary from the radius of convergence in Fig. \ref{fig:hl10} is
    \begin{equation}
        S_{*}^{-1} = 0.0236 \implies \abs{h/c} \sim 0.348
    \end{equation}

    \item $h_{L} = 100$

    \begin{figure}[h!]
        \centering
        \includegraphics[width=9.5cm]{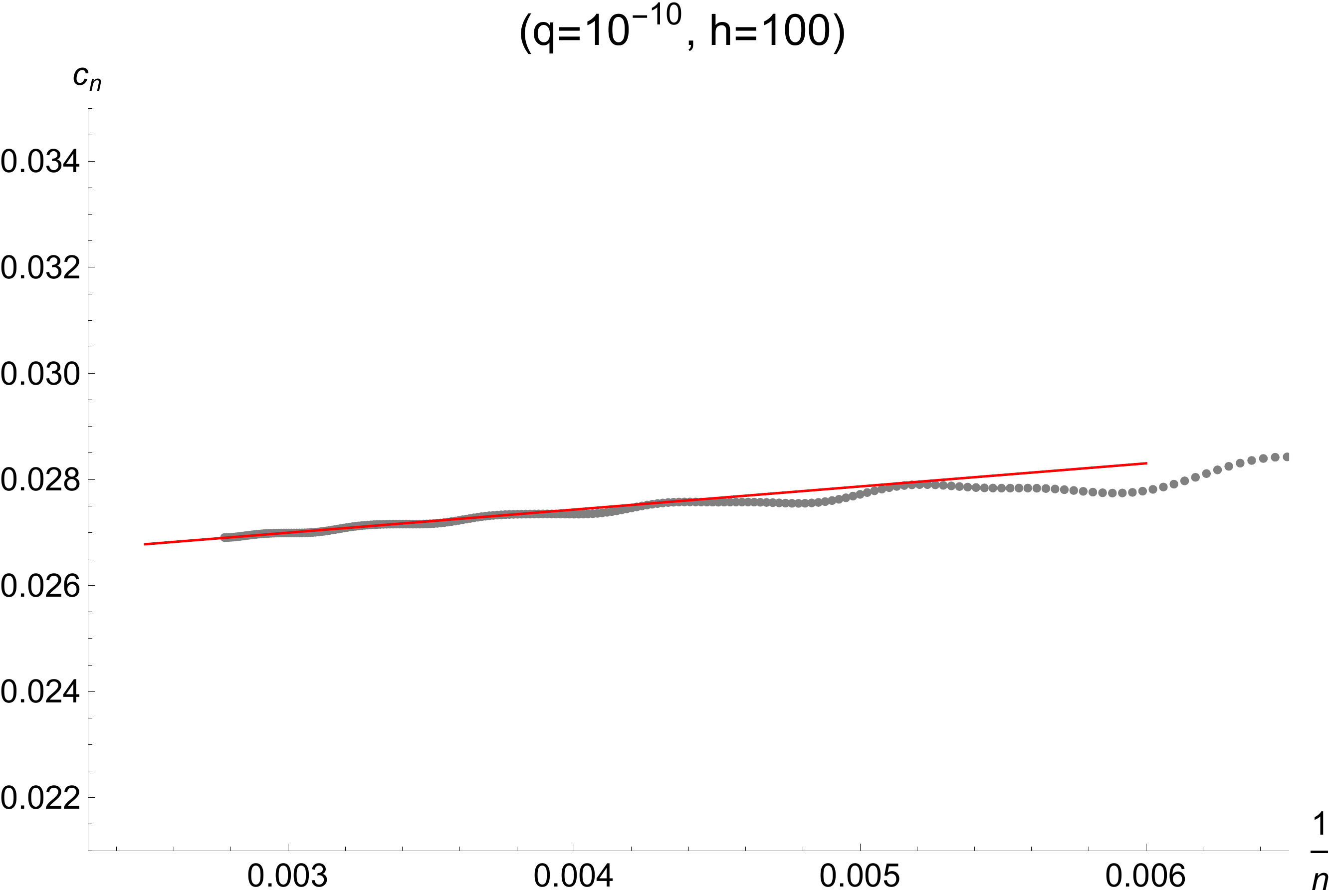} 
        \hspace{1cm} 
        \caption{$c_{n}$ vs $1/n$ for $h_{L}=100$}
        \label{fig:hl100}
    \end{figure}

    The value of the intercept and the dimension of the primary from the radius of convergence in Fig. \ref{fig:hl100} is
    \begin{equation}
        S_{*}^{-1} = 0.0257  \implies \abs{h/c} \sim 0.320
    \end{equation}

\end{itemize}

\subsection{All-heavy block $\langle \phi_{h_{H}}\phi_{h_{H}}\phi_{h_{H}}\phi_{h_{H}} \rangle$}

\begin{itemize}
    \item $h_{H} = c/9$
    \begin{figure}[h!]
        \centering
        \includegraphics[width=9.5cm]{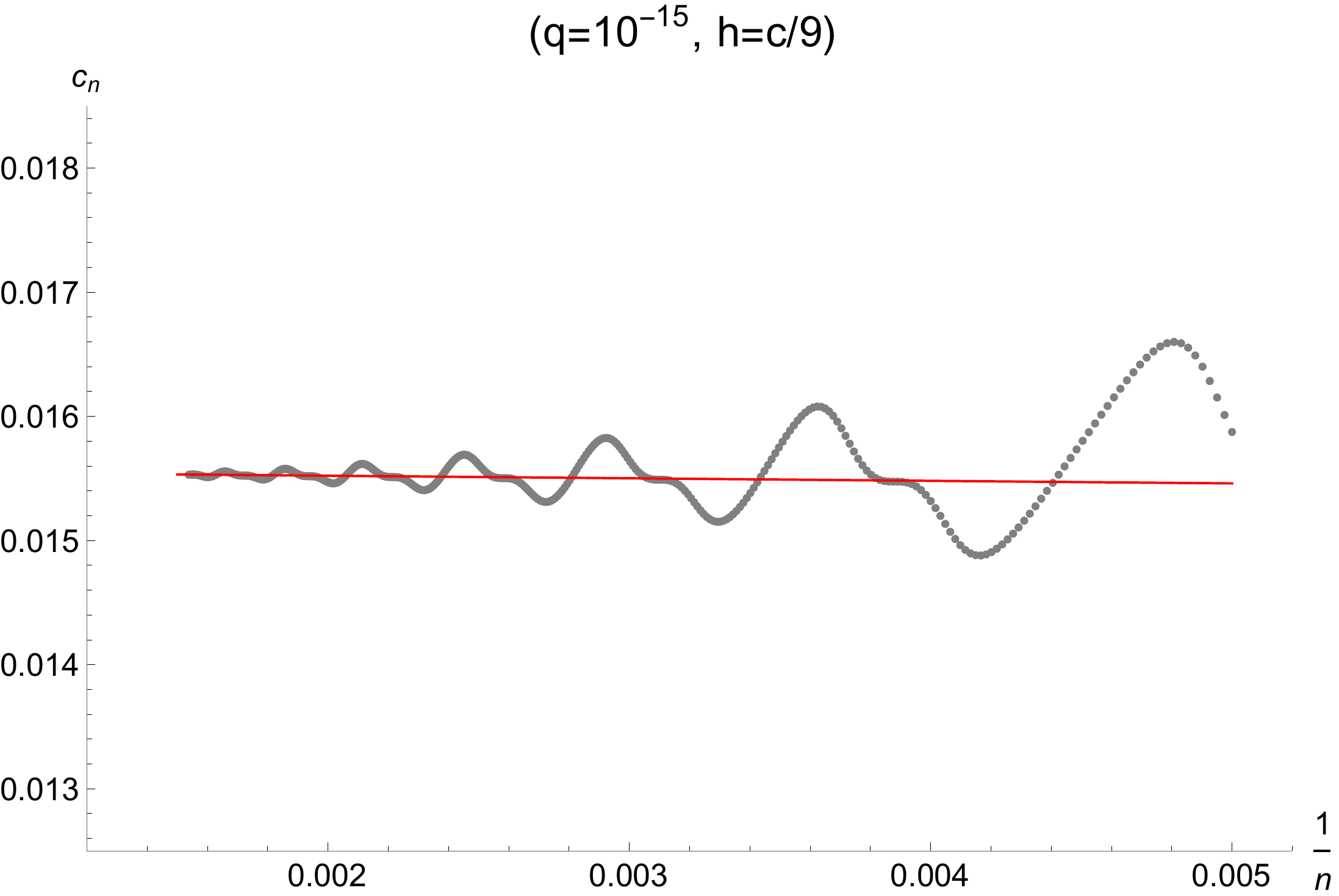} 
        \hspace{1cm} 
        \caption{$c_{n}$ vs $1/n$ for $h_{H}=c/9$.}
        \label{fig:c9heavy}
    \end{figure}

    The value of the intercept and the dimension of the primary from the radius of convergence in Fig. \ref{fig:c9heavy} is
    \begin{equation}
        S_{*}^{-1} = 0.0156  \implies \abs{h/c} \sim 0.337
    \end{equation}

    \item $h_{H} = c/12$
    \begin{figure}[h!]
        \centering
        \includegraphics[width=9.5cm]{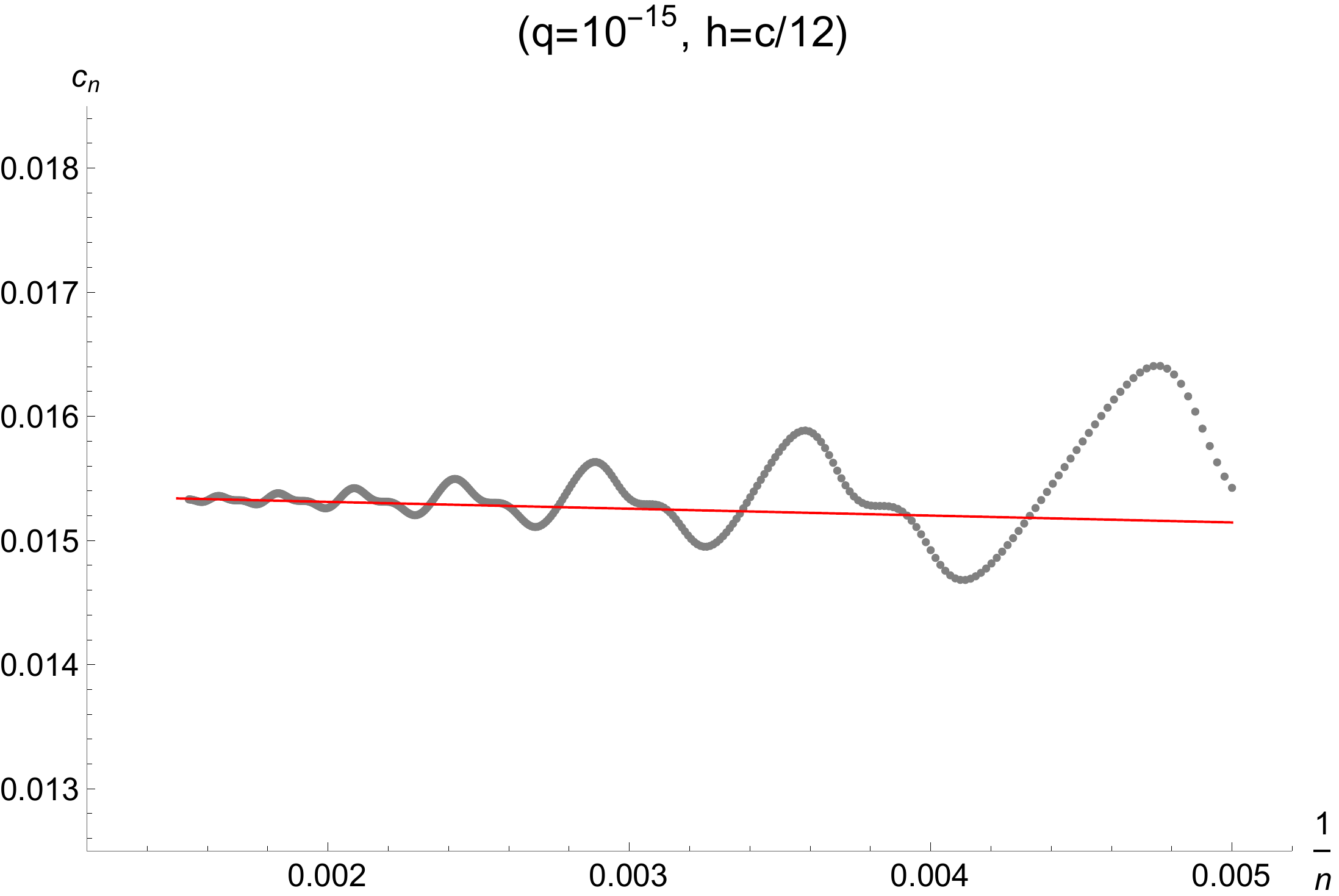} 
        \hspace{1cm} 
        \caption{$c_{n}$ vs $1/n$ for $h_{H}=c/12$.}
        \label{fig:c12heavy}
    \end{figure}

    The value of the intercept and the dimension of the primary from the radius of convergence in Fig. \ref{fig:c12heavy} is
    \begin{equation}
        S_{*}^{-1} = 0.0154  \implies \abs{h/c} \sim 0.340
    \end{equation}

\end{itemize}

\section{Euler class}
\label{app:eulerclassstuff}
In this appendix we describe more explicitly how the components of $PSL(2,{\mathbb R})$ moduli space with Euler class $|e|<2g-2$ correspond to geometries with integer conical excesses.  We will do so by focusing on the simplest example, that of $PSL(2,{\mathbb R})$ connections in empty AdS${}_3$, which is topologically $\Sigma\times{\mathbb R}$ where $\Sigma$ is a disk.  

Starting with an $SL(2,\mathbb{R})$ gauge field $A=A^{a}t^{a}$, the Euler class is defined by restricting this gauge field to its $SO(2)$ (or $U(1)$) component. Although the full $F=dA + A\wedge A$ is flat, the curvature $F^{U(1)}=dA^{U(1)}$ of this component of the gauge field is non-zero.  The Euler class is the integral of this curvature 
\begin{equation}
e= \frac{1}{2\pi}\int_{\Sigma} F^{U(1)}
\end{equation}
over the spatial slice $\Sigma$.
We will now compute this integral when $\Sigma$ is the hyperbolic disk, the constant time slice of AdS$_3$.

We will take the generators of $sl(2,\mathbb{R})$ to be
\begin{equation}
    t^{0} = i\frac{\sigma_{y}}{2} \,,\quad t^{1} = \frac{\sigma_{x}}{2} \,,\quad t^{2} = \frac{\sigma_{z}}{2}
\end{equation}
so that 
\begin{equation}
    [t^{0},t^{1}] = t^{2} \,, \quad [t^{1},t^{2}] = -t^{0} \,,\quad  [t^{2},t^{0}] = t^{1}
\end{equation}
\begin{equation}
    2tr(t^{a}t^{b}) = \eta^{ab} \,,\quad \eta = \diag(-1,1,1,1)
\end{equation}
Then $t^0$ is the generator of $so(2)$, so $A^{U(1)}=A^{0}$ and
\begin{equation}
    F^{U(1)} = dA^{0}|_{t=0}
\end{equation}

We consider AdS$_{3}$ in global coordinates (and units $\ell_{AdS}=1$), with 
\begin{equation}
    ds^{2} = -\cosh^{2}(\rho)\, dt^{2} + d\rho^2 + \sinh^2(\rho)\,d\phi^2
\end{equation}
so that
\begin{equation}
    \label{7}
    A^0 = \cosh\rho (dt + d\phi) \,,\quad A^1 = \sinh\rho (dt + d\phi) \,,\quad A^2 = d\rho
\end{equation}
where $A=\omega-e$, with 
similar expressions for $\bar{A}$. The curvature of the $U(1)$ component is 
\begin{equation}
    F^{U(1)} = -\sinh\rho\, \left(d\phi \wedge d\rho+dt\wedge d\rho\right)
\end{equation}
so that
\begin{equation}
    \frac{1}{2\pi}\int_{\Sigma} F^{U(1)} = -\frac{1}{2\pi}\int_{0}^{2\pi} d\phi \int_{0}^{R} \sinh\rho = 1-\cosh R~.
\end{equation}
Here we have introduced a radial cutoff at $\rho=R$ to regulate the integral. A simple counter-term which removes this divergence is 
\begin{equation}
    \frac{1}{2\pi}\oint_{\rho=R} A^0|_{t=0} = \frac{1}{2\pi} \int_{0}^{2\pi} d\phi \cosh R = \cosh R
\end{equation}
The result is that the ``regulated" Euler class
\begin{equation}
    e(\Sigma) := \frac{1}{2\pi}\int_{\Sigma} F^{U(1)} + \frac{1}{2\pi}\oint_{\rho=R} A^0|_{t=0}
\end{equation}
is independent of $R$.  We conclude that in AdS$_3$ we have 
\begin{equation}
    e(\text{disk}) = 1
\end{equation}
Of course, this is just the usual Euler characteristic of the disk.

We can now understand how this is modified when we introduce a conical singularity at the origin. The metric is the same, except now 
\begin{equation}
    \phi \sim \phi + 2\pi(1-\eta)
\end{equation}
where $2\pi\eta$ is the conical defect/excess.
Following the previous steps  gives
\begin{equation}
    \frac{1}{2\pi}\int_{\Sigma} F^{U(1)} = -\frac{1}{2\pi}\int_{0}^{2\pi(1-\eta)} d\phi \int_{0}^{R} \sinh\rho = (1-\eta)(1-\cosh R)
\end{equation}
\begin{equation}
    \frac{1}{2\pi}\oint_{\rho=R} A^0|_{t=0} = \frac{1}{2\pi} \int_{0}^{2\pi(1-\eta)} d\phi \cosh R = (1-\eta)\cosh R
\end{equation}
So the Euler class of AdS$_3$ with a conical defect is
\begin{equation}
    e(\text{disk}_{\eta}) = 1-\eta
\end{equation}
We see that the difference between the Euler class for the above two cases is 
\begin{equation}
    e(\text{disk}) - e(\text{disk}_{\eta}) = \eta
\end{equation}
Our conclusion is that the introduction of a conical excess of angle $2\pi n$ simply shifts the Euler class by an integer.  This was exactly the situation encountered in section \ref{sec:csquant}, where we saw that the non-Teichm\"uller components of $PSL(2,{\mathbb R})$ moduli space have Euler class which differs from that of the Teichm\"uller component by an integer.

\bibliography{References}
\bibliographystyle{JHEP}

\end{document}